\documentclass[11pt,preprint]{aastex}
\usepackage{colordvi}
\usepackage[usenames]{color}   

\newcommand{\mgamma}{\gamma_{\rm eff}}
\def\ltw{\>\hbox{\lower.25em\hbox{$\buildrel<\over\sim$}}\>}
\def\gtw{\>\hbox{\lower.25em\hbox{$\buildrel>\over\sim$}}\>}

\slugcomment{to appear in The Astrophysical Journal}

\shortauthors{Hardee \& Eilek}
\shorttitle{Twisted Filaments in the M\,87 Jet}

\begin{document}
\baselineskip 12 pt
\parskip 4pt

\title{Using Twisted Filaments to Model the Inner Jet in M\,87}

\author{P. E. Hardee\altaffilmark{1} and J. A. Eilek\altaffilmark{2,3}}
\altaffiltext{1}{Department of Physics \& Astronomy, The University of Alabama,
Tuscaloosa, AL 35487}
\altaffiltext{2}{Physics Department, New Mexico Tech, Socorro, NM 87801}
\altaffiltext{3}{National Radio Astronomy Observatory, Socorro NM 87801.  
NRAO is a facility of the National Science
Foundation, operated under cooperative agreement by Associated
Universities, Inc.}
\email{phardee@bama.ua.edu}

\begin{abstract}
\baselineskip 12 pt
Radio and optical images of the M\,87 jet show bright filaments, twisted into an apparent double helix, extending from HST-1 to knot A.  Proper motions within the jet suggest a decelerating jet flow passing through a slower, accelerating wave pattern.  We use these observations to develop a mass and energy flux conserving model describing the jet flow and conditions along the jet.  Our model requires the jet  to be an internally hot, but subrelativistic plasma, from HST-1 to knot A. Subsequently we assume that the jet is in pressure balance with an external cocoon and we determine the cocoon conditions required if the twisted filaments are the result of the Kelvin-Helmholtz (KH) unstable elliptical mode.  We find that the cocoon must be cooler than the jet at HST-1 but must be about as hot as the jet at knot A. Under these conditions we find that the observed filament wavelength is near the elliptical mode maximum growth rate and growth is rapid enough for the filaments to develop and saturate well before HST-1.  We generate a pseudo-synchrotron image of a model jet carrying a combination of normal modes of the KH instability.  The pseudo-synchrotron image of the jet reveals: (1) that a slow decline in the model jet's surface brightness is still about five times faster than the real jet; (2) that KH produced dual helically twisted filaments can appear qualitatively similar to those on the real jet if any helical perturbation to the jet is very small or nonexistent inside knot A; (3) that the knots in the real jet cannot be associated with the twisted filamentary features and are unlikely to be the result of a KH instability. The existence of the knots in the real jet, the limb brightening of the real jet in the radio, and the slower decline of the surface brightness of the real jet indicate that additional processes --- such as unsteady jet flow and internal particle acceleration --- are occurring within the jet. Disruption of the real jet beyond knot A by KH instability is consistent with the jet and cocoon conditions we find at knot A.

\end{abstract}

\keywords{galaxies: individual (M87) --- galaxies: jets --- 
galaxies: active --- radio continuum: galaxies --- hydrodynamics --- 
relativity \vspace{-0.6cm}}

\section{Introduction}

M\,87 (Virgo A, NGC\,4486, 3C\,274) is a giant elliptical galaxy near the center of the Virgo Cluster.  The well-known, kpc-scale jet in this galaxy is a prominent source of radio, optical, near-ultraviolet (NUV) and X-ray emission. There are remarkable similarities between the radio, optical and NUV emission on scales $\sim 0.1\!-\!1$ kpc. The bright knots, the filamentary emission between the knots, and the bends and twists in the jet
can easily be identified in the optical and NUV as well as the radio.  While X-ray images from Chandra are not of comparable resolution to the radio or optical, the same bright knots and overall jet structure can easily be seen in the X-rays. Relativistic proper motions are seen within the knots, both in the radio and the optical. 

A variety of models have been proposed for the physical state of the kpc-scale jet, but  none has emerged as definitive.  \citet{BB96} proposed a hydrodynamic model, in which the knots are shocks caused by the Kelvin-Helmholtz (KH) instability. \citet{HB97} extended the hydrodynamic model, arguing that intensity changes and spectral evolution along the jet can be explained by shock compression of a weakly magnetized plasma. Other authors have put forth  models in which the magnetic field plays a key role. \citet{FB1989} and also \citet{WMB1994}  proposed that the jet is a force-free  MHD configuration (following \citet{KC1985}).  In this scenario the knots correspond to ``magnetic islands'' in the force-free state. More recently, \citet{GTB2005},  also \citet{G2009}, presented an MHD model of the jet's launching from the central black hole;  they identified HST-1 as a recollimation shock.

In this paper we return to hydrodynamic models and consider the pair of twisted emission filaments reported by Lobanov, Hardee \& Eilek (2003; hereafter LHE).  We propose that these filaments are created by the KH
instability in a weakly magnetized, predominantly hydrodynamic flow. Several authors have used the properties of KH modes to estimate the physical conditions in jets and their surroundings (e.g., Perucho et al.\ 2004). The method has been applied to knots in the the M\,87 jet \citep{BB96}; to  twisted structures in the  3C\,120 jet \citep{HWG2005};  to twisted emission threads in the 3C\,273 jet \citep{PLMH06}, to superluminal motions and accelerations of components along curved trajectories in the 3C\,345 jet \citep{H87,S95}, and to transverse jet structure in 0836+710 \citep{PL07}.  

We base our study of the M\,87 jet on a linear analysis of the KH instability, e.g., \citet{H00,H07}, supplemented by our experience with numerical simulations of its development and saturation.  We chose this approach because our goal is to constrain the physical conditions in the M\,87 jet and its surroundings, and it would be prohibitively expensive to run the large suite of simulations that would be necessary for a good exploration of parameter space.  For example, computational constraints make it difficult to simulate  an expanding, three-dimensional, relativistic jet, close to the line of sight, over the distance from HST-1 to knot A, with sufficient spatial resolution and temporal storage.  Such a simulation would need to have enough computational zones to satisfy projection and numerical viscosity issues \citep{P04}.  The simulation would also have to span and store enough time information to deal with the light-travel time effects that lead to the observed superluminal motions \citep{G97, A03} and that make the apparent structure different from the intrinsic structure.  However, more recently \citet{Mimetal2009} have developed a Lagrangian algorithm that accomplishes this task in a better and more physically correct way.

We proceed and organize the paper as follows. In Section \ref{flow_obs} we review the structure of the observed jet, particularly the filaments identified by LHE, and the proper motions of the bright knots.  We use the proper motions to estimate the flow speed of the jet plasma and the wave speed of the KH pattern. 
The data suggest that the jet decelerates between HST-1 and knot A; this is a key point of our models.  In Section \ref{flow_models} we develop an energy-conserving jet model which matches the observed flow deceleration and determines the thermal state of the jet plasma. We also develop  models in which a small part of the energy flux is lost to the cocoon.  In both models we assume the jet is in pressure balance with an external ``cocoon'', as required by our KH analysis.   In Section \ref{KH_Modes} we present a KH instability analysis, and determine what the thermal state of the cocoon must be if the elliptical mode of the  KH instability causes  the observed filaments.  We do this by matching solutions of the wave dispersion relation to observations of the pattern speed and wavelength along the jet.  We also verify that the KH instability growth rate under these conditions is sufficiently rapid for the instability to saturate. In Section \ref{Scaling} we explore the likely internal state of the jet and cocoon plasmas.   In Section \ref{Pseudo_Synch} we present a pseudo-synchrotron image of the jet, to compare our most likely model to the observed intensity structure of the real jet.  Finally, in Section \ref{Summary} we briefly summarize our  results and in Section \ref{Discussion} we discuss a few implications of our models.

\vspace{-0.70cm} 
\section{Jet Structure and Environment}
\label{flow_obs}
\vspace{-0.10cm} 

In this section we review the basic structure and immediate environment of the M\,87 jet.  We also review low-order wave modes created by the KH instability and compare them to the structure of the real jet. Because our KH analysis requires a model of the jet flow speed as well as the speed of the wave pattern within the jet, we use observed proper motions to determine the evolution of both speeds along the jet. 

\begin{figure}[h!]
\vspace{-5.60cm}
{\center 
\includegraphics[width=0.9\textwidth]{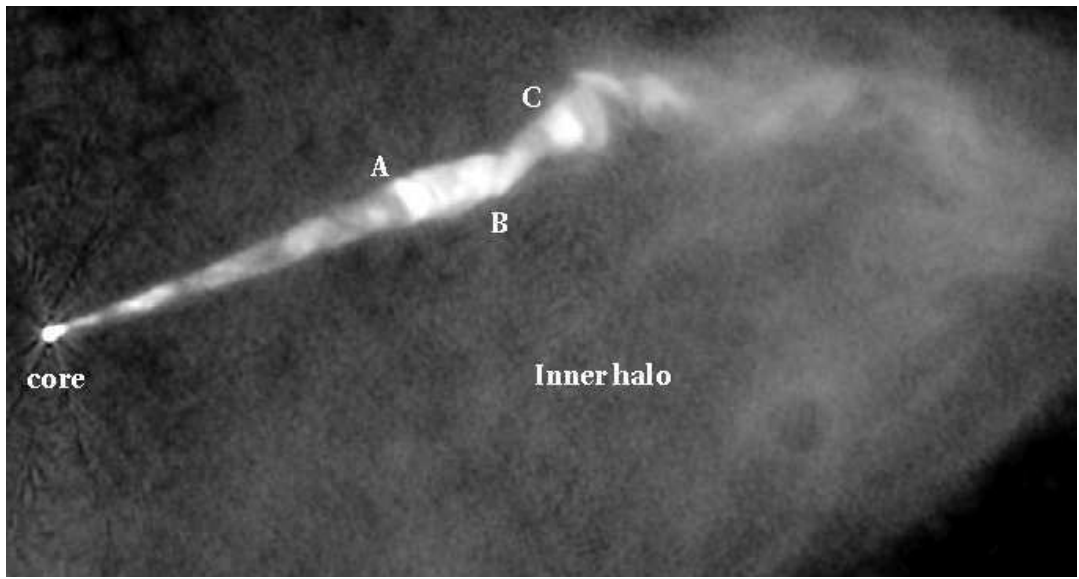}
\vspace{-5.80cm}
\caption{\footnotesize \baselineskip 11pt Radio image of the jet and western part of the inner radio halo in M\,87. The three bright knots A, B, C are labelled, as is the western part of the inner radio halo. The straight, well-collimated inner jet, which we model, extends from the core to knot A.  The jet begins to destabilize past knot A, and disrupts fully past knot C, where the plasma flow bends and feeds the inner halo.   VLA image taken at 15 GHz, with resolution 0.15 arcsec, from OHC.}
\label{Jet_large}}
\end{figure}

The overall structure of the M\,87 jet as seen in the radio is shown in Figure \ref{Jet_large} (from Owen, Hardee \& Cornwell (1989); hereafter OHC). The brightest emission knots in the outer part of the jet  are labelled A, B and C.  The inner jet --- from the core to knot A ---  is expanding uniformly,  limb brightened with sharp edges.  Bright knots and fainter filaments can be seen within the flow.  Past  knot A the jet brightens, stops expanding, and maintains a more uniform radius. Knots B and C contain tightly twisted filaments as well as some bright emission patches.  Past knot A the flow develops significant sideways motion, and past knot C it breaks up dramatically.   
We take the distance to M\,87 to be $16.7$~Mpc \citep{Mei07}, thus $1\arcsec \sim 80$ pc.  As we discuss below, the jet probably lies between $15\arcdeg$ and $25\arcdeg$ from the line of sight, so its intrinsic length is $\sim 2.4\!-\!3.9$ times longer than its apparent (projected) length.

The plasma content of the jet remains unknown.  Because the jet is a synchrotron source, we know it must contain a relativistic, magnetized lepton plasma.  \citet{RFCR1996}  argued for a pure lepton jet, because a cold, electron-ion jet would violate constraints on the opacity and surface brightness of the self-absorbed, pc-scale core. However, more recent evidence on the jet viewing angle (as in Section \ref{Jet_speed}) and the internal energy of the jet \citep{BB96} and our results in Section \ref{Jet_structure_models} expand the possible parameter space to allow electron-ion jets within the arguments of \citet{RFCR1996}.

The local environment of the jet, on projected scales $\ltw 2$ kpc, is well studied in radio and X-rays. 
 From the radio, we know that the jet feeds, and appears to be surrounded by,  inner radio lobes which extend  $\sim 3$ kpc from the galactic core.  The western part of the inner lobes can be seen in Figure \ref{Jet_large}.  The inner radio lobes contain a relativistic, magnetized lepton plasma. The minimum pressure allowed by the
synchrotron power of the inner lobes is below that of the local interstellar medium (ISM) throughout most of the lobes, but a few bright filaments are at higher pressure \citep{HOE1989}.  Such bright filaments are probably
transient, suggesting jet-driven turbulence in  the radio lobes. 

From the X-rays, we know that the jet sits within the central ISM of M\,87.  That ISM is an X-ray loud, ion-electron plasma; averaged over the inner few kpc it has $T \sim 10^7$K and $n \sim 0.1$~cm$^{-3}$,
e.g., \citet{NB1995}, also \citet{OEK2000}. The inner  ISM is magnetized; \citet{OEK1990}  detected Faraday rotation from the ISM in front of the jet and inner radio lobes. The inner ISM is  disordered and turbulent;  filaments, loops and bubbles are apparent in the inner few kpc of the Chandra image \citep{Form05,Form07}.  The correlation of features in the X-ray image with those in the radio halo shows  that the radio-loud plasma from the jet is interacting, and probably mixing, with the thermal ISM. \citet{KOE1996}  found that emission-line clouds in the inner few kpc of M\,87  show disordered motion at about half the sound speed of the local ISM;  these clouds probably share the velocity field of the ISM.

\vspace{-0.7cm} 
\subsection{Twisted Filaments in the M\,87 Jet}
\vspace{-0.1cm}

\begin{figure}[h!]
\vspace{-6.15cm}
{\center
\includegraphics[width=0.9\textwidth]{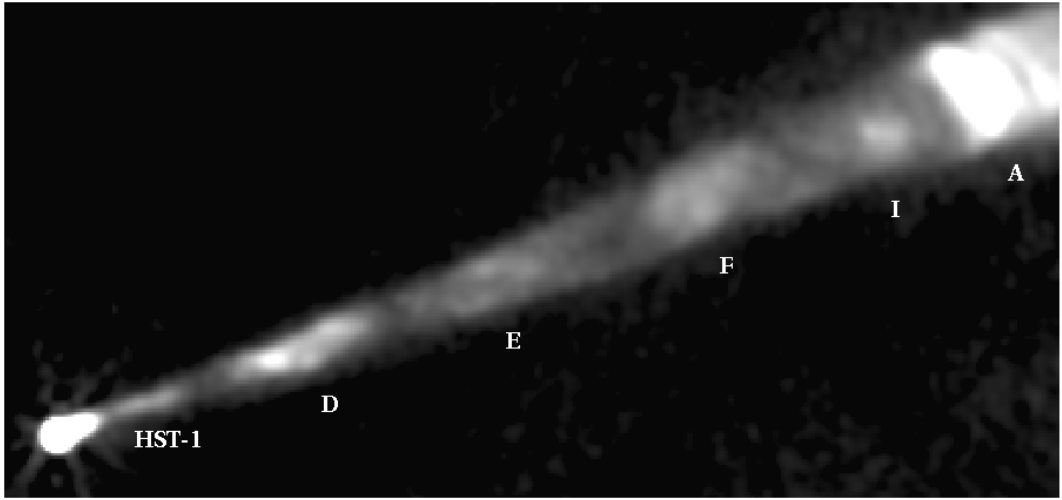}
\vspace{-6.15cm}
\caption{\footnotesize  \baselineskip 11pt The inner part of the M\,87 jet, from the same image as shown in Fig \ref{Jet_large}, displayed with a logarithmic response to enhance regions of low surface brightness.  These VLA observations were carried out in 1985, when knot HST-1 was in a  quiescent, pre-flare state, enabling the surrounding jet structure to be imaged more clearly. In addition to the bright knots HST-1, D, E, F, I and A, this image shows the twisted, helical filaments between the knots which LHE studied. }
\label{Jet_inner}}
\end{figure}

Figure \ref{Jet_inner} shows the inner jet in M\,87.  The prominent emission knots are labelled, from the core 
outwards, as HST-1 (formerly called knot G), D, E, F, I and A.  HST-1 is $\sim 0.9\arcsec$ (projected distance $\sim 70$ pc) from the core; knot A is  $\sim 12\arcsec$ 
(projected distance $\sim 1$ kpc) from the core.  The jet expands with a constant apparent
opening angle, $\sim 6\arcdeg$, from well inside HST-1 to knot A. Between HST-1 and knot A the emission profile is edge-brightened, relatively sharp in the radio, and contains the twisted filaments reported by LHE. The projected magnetic field vectors lie more or less along the edge of the jet flow (OHC), which may be indicative of a
shear layer \citep{L1981}. However, there is no indication of mass entrainment into a turbulent surface layer as suggested for 3C\,31 on large scales \citep{LB2002}. 

The jet structure seen in the optical and NUV  is quite similar  to that seen in radio, but there are important differences in  detail.  The optical/NUV emission is more concentrated in the knots  and towards the jet axis, and the outer edges of the jet are less   well defined than in the radio e.g., \citet{SBM1996} or   \citet{Madrid2007}.  The optical/NUV knots also differ somewhat in  polarization from the radio knots \citep{P1999}.  The major bright
  radio or optical knots can also be identified in the  lower-resolution X-ray image, e.g., \citet{PW2005}, but they can  differ in position or structure from their radio or optical   counterparts.

In this paper we do not focus on the knots, but rather focus on the  twisted filaments which LHE detected in the inner jet of  M\,87. LHE extracted slices transverse to the overall  jet axis, spaced by the pixel size, in radio (VLA) and optical (HST)  images.  They fit each slice with double Gaussian profiles, allowing  the amplitude, position and width of the two Gaussians to vary at  each slice.  Both VLA and HST images revealed a consistent pattern:  the jet contains two intertwined emission filaments, which can be  traced from HST-1 out to knot A.  Unlike the situation with the knots,  LHE found no significant differences between the filaments in the VLA and HST image. LHE interpreted these results as  evidence for a twisted-helix structure, wrapped around the edge of
  the jet and emitting in radio and optical bands, between HST-1 and knot A. The observed radial difference in filament and knot location within the jet suggests that we can model the filaments without considering the mechanism responsible for the knots. 

Looking ahead to our KH analysis, the most important result from LHE is that the observed filament wavelength (projected on   the sky) is not constant; it increases from $\lambda_p^{\rm ob} \sim  2\arcsec \pm 0.3\arcsec$ at HST-1 to $\lambda_p^{\rm ob} \sim  3\arcsec \pm 0.3\arcsec$ at knot A.  We also point out that the bright knots in the inner jet do not, in general, coincide with filament crossings. Although knot E appears to coincide with a filament crossing, Figure \ref{Jet_inner} shows that knots D, F, I and A are more complex; this is true in the optical as well as the radio images.

\vspace{-0.7cm} 
\subsection{Twisted Filaments and the KH Instability}
\label{KH_intro}
\vspace{-0.1cm}

In what follows, we determine the implications for the jet and its immediate environment if the filaments are generated by KH instability. The KH instability can occur when velocity shear exists across a fluid
boundary, such as the jet surface.  It manifests itself in the form of pressure waves generated at or near the jet surface.  The pressure waves can resemble the twisted filaments in the M\,87 jet. Small perturbations to the  jet flow can be described mathematically in terms of a sum of Fourier components, referred to as
``normal modes'' (as in Section \ref{KH_Modes} and Appendix A).  The two lowest-order twisting
  modes, helical and elliptical, describe associated distortions to the jet surface.  Although higher-order modes initially grow more rapidly, the lower-order modes dominate at nonlinear levels, e.g., \citet{HCR97}.
Each mode involves both surface waves and body waves, which have different radial structures and move at different speeds.  We will show (in Section \ref{growth_rates}) that surface waves dominate in the M\,87 situation.

\begin{figure}[h!]
\vspace{-3.0cm}
{\center 
\includegraphics[width=1.0\textwidth]{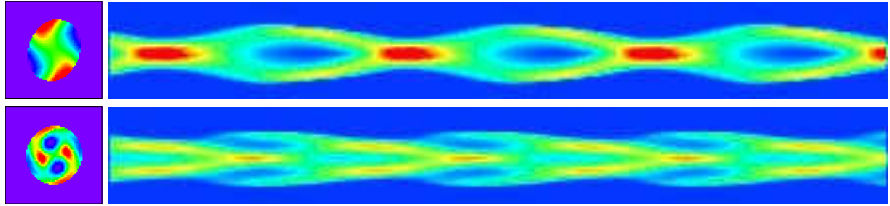}
\vspace{-16.250cm}
\caption{\footnotesize \baselineskip 11pt  Generic pressure cross sections (left) and line-of-sight pressure integrations  (right) for elliptical surface (top) and elliptical first body (bottom) modes of a jet lying in the plane of the sky.  Red indicates high and blue indicates low pressure. No relativistic or light-travel effects are included in the line-of-sight images.} 
\label{illustrate_KH}}
\end{figure}

In this paper we argue that the elliptical mode is dominant in the inner jet,   and gives rise to the filaments described by LHE. We illustrate the pressure structure of the surface and first body elliptical modes in Figure \ref{illustrate_KH}, which shows a typical cross section and an integrated line of sight pressure image for each
mode \citep{H00}. The elliptical surface mode generates two high-pressure filaments near the jet surface.  The first body mode generates two high-pressure filaments located at about the radius midpoint, as well as a spiral, high-pressure pattern between these filaments and the jet surface.
Because pressure can be used as a simple proxy for synchrotron emissivity (see Section \ref{Pseudo_Synch}), Figure \ref{illustrate_KH} illustrates how the the surface and body elliptical modes might appear in a radio jet lying in the plane of the sky.  Comparison of Figures \ref{Jet_inner} and \ref{illustrate_KH}, along with
the double Gaussian fits from LHE, indicate that the filaments in the M\,87 jet lie near to the jet surface and resemble the elliptical surface mode.

We thus suggest that the elliptical surface mode dominates in the inner M\,87 jet.  Relativistic hydrodynamic simulations  \citep{HCR97,HH03} suggest that the elliptical surface mode   saturates without disrupting the jet, perhaps because this mode does   not displace the jet's centerline.  We note that the elliptical body
  mode grows more slowly than the surface mode in a linear analysis   (see Section \ref{growth_rates}), and simulations show the body  modes saturate at lower amplitudes than do the surface modes  \citep{H01, HH03}.

Our hypothesis that elliptical modes dominate in the inner M\,87 jet disagrees  with \citet{BB96} who suggested the bright knots in the M\,87 jet  are caused by the helical mode. In particular, the inner M\,87 jet
  between HST-1 and knot A (as in Figure \ref{Jet_inner}) does not resemble  the morphology of the helical mode, e.g., \citet{H00}.  The helical  mode displaces the jet centroid, causing a lateral oscillation in  the jet; but the inner M\,87 jet has very straight edges, with no  sign of lateral displacement.  In the absence of significant lateral oscillation,  the helical surface mode could still appear as a  single, high-pressure filament wrapped around the jet; this disagrees with  the dual, double-helix filaments that exist in the real inner  jet. By comparison, the elliptical mode does not displace the jet  centroid, but it does create a pair of high-pressure twisted  filaments which wrap around the jet.  Furthermore, no helical  twisting has been detected at sub-parsec scales \citep{Acc09};
  this suggests there is only a minimal helical perturbation  to the jet inside HST-1.

The apparent absence of a KH helical mode is perhaps surprising.  Linear analysis (as in Section \ref{growth_rates}) finds the helical  mode will grow more slowly than higher-order modes in the M\,87 jet,
  but still rapidly enough that one might expect it to be detectable.  Simulations agree: \citet{HCR97}, also \citet{HH03}, find that the  helical mode, if driven, grows slowly but surely, and does not
  saturate.  Instead, it continues to grow until it reaches nonlinear  levels and disrupts the jet.  Thus, the lack of any detectable  helical mode, either on sub-parsec scales or between HST-1 and knot  A, suggests that any initial helical-mode perturbation to the jet  must be very small.

\vspace{-0.7cm} 
\subsection{Jet Flow and Pattern Motions}
\label{Jet_speed}
\vspace{-0.1cm}

In order to use the observed filament wavelengths to determine conditions in the jet and its surroundings, we must determine the viewing angle, the jet flow speed and the speed of the wave pattern within the jet. To this end, we take advantage of the extensive data on proper motions of radio and optical features in the jet.  Because
we expect the KH instability to give jet flow {\it through} wave patterns, we identify faster proper motions with the jet flow, and slower proper motions with a pattern speed. The similarity of the optical and radio jet filaments, and the lack of evidence for a broad velocity shear layer, make it likely that we can use the fastest observed proper motions from both bands to reliably represent the jet speed.  In what follows, we use proper motion data to develop continuous, steady-state models of the jet flow and pattern speed.  By doing this, we implicitly assume that whatever causes the emission enhancement in the bright knots does not create any significant deviations from smooth flow within the jet.

\vspace{-0.8cm} 
\subsubsection{Viewing Angle}
\vspace{-0.1cm}

To constrain the viewing angle, we look to the fastest observed proper motions. The fastest optical proper motion implies a superluminal speed $\beta^{\rm ob} \sim 6.1 \pm 0.6$ at the position of HST-1 \citep{BSM99}.  The most rapid radio proper motion implies $\beta^{\rm ob} \sim 4.3 \pm 0.7$ \citep{CHS07} at the position of HST-1. Recalling that the maximum viewing angle is given by $\cos\theta_{\rm max} = [(\beta^{\rm ob})^2 - 1]/[(\beta^{\rm ob})^2 + 1] $, the highest optically determined superluminal speed requires a jet viewing angle $\theta < 18.6\arcdeg$.  On the other hand, the highest radio-determined superluminal speed requires only $\theta < 26.1\arcdeg$.  We therefore work with two models, jets at $15\arcdeg$ and $25\arcdeg$ to the line
of sight, taking these as representative of limits on the jet viewing angle and the jet speed at HST-1.

\vspace{-0.8cm} 
\subsubsection{Jet flow speed}
\vspace{-0.1cm}

To constrain the flow speed of the jet plasma, we again look to the fastest observed proper motions. The fastest optical proper motions decline along the jet with superluminal speeds $\beta^{\rm ob} \sim 6.1 \pm 0.6, \sim 5.11 \pm 0.66, \sim 5.1 \pm 0.9$, and $\sim 3.9 \pm 0.8$ found at the positions of HST-1, HST-2, knot D, and knot E,
respectively \citep{BSM99}. The fastest radio proper motions also decline along the jet with speeds of $\beta^{\rm ob} \sim 4.3 \pm 0.7$ at HST-1 \citep{CHS07} and $\beta^{\rm ob} \sim 2.5 \pm 0.3$ at knot D \citep{BZO95}. We use the radio proper motion downstream of knot A, $\beta^{\rm ob} \sim 1.25 \pm 0.13$ at knot B \citep{BZO95}, to set a lower limit on the flow speed at knot A.  For  HST-1 and knot D we observe flow through a more slowly moving knot. Thus, the fastest observed radio and optical proper motions associated with the knots suggest a flow
speed which {\it decreases} along the jet between HST-1 and knot A. At a viewing angle of $15\arcdeg$, the four optical data points spanning the $6\arcsec$ from HST-1 to knot E require more than a $34\%$ decrease in the Lorentz factor and provide the most robust evidence for a decreasing jet speed.

We synthesize these data into two jet models. To represent the faster optical  motions and a smaller
viewing angle we work with a {\it fast jet model} with viewing angle $\theta = 15\arcdeg$ and intrinsic jet full opening angle $2\psi = 0.0272$~radian~$ = 1.59\arcdeg$.  The jet Lorentz factor is $\gamma_{j0} = 7.50$ ($\beta^{\rm ob} \sim 6$) at HST-1,  $\gamma_{j} \sim 5.40$ at knot D and $\gamma_{j} \sim 4.15$ at knot E.  To represent the slower radio motions and a larger viewing angle we work with a {\it slow jet model} with $\theta = 25\arcdeg$, full opening angle $2\psi = 0.0444$~radian~$= 2.54\arcdeg$. In this model, the jet Lorentz factor is 
$\gamma_{j0} = 4.40$ ($\beta^{\rm ob} \sim 3.5$) at HST-1, and $\gamma_{j} \sim 3.10$ at knot D. Our fast and slow jet models are constrained by a lower limit to the jet Lorentz factor at knot A,  $\gamma_j \ge 1.90$ for $\theta = 15\arcdeg$ and $\gamma_j \ge 1.70$ for $\theta = 25\arcdeg$. We return to these two jet  models in Section \ref{flow_models}. 

\vspace{-0.8cm} 
\subsubsection{Observed pattern speed}  
\vspace{-0.1cm}

To constrain the pattern speed, we turn to slower, subluminal, optical or radio proper motions. Here our best
evidence comes from the slower speeds associated with HST-1 or knot D, because these two knots provide clear evidence for jet flow through more slowly moving structures (patterns), as might be expected for the
KH instability.  While knot D shows a broad range of motions, the average radio-determined speed for knot D is $\beta^{\rm ob} \sim 0.41 \pm 0.10$ \citep{BZO95}.  The story at HST-1 is more complex.  The upstream edge of HST-1 has been measured to move at $\beta^{\rm ob} \sim 0.84 \pm 0.11$ (\citet{BSM99}, optical), at $\beta^{\rm ob} < 0.25$ (\citet{CHS07}, radio), and at $\beta^{\rm ob} \sim 0.61 \pm 0.31$ (\citet{CRKL2010}, radio). In addition to the known variability of HST-1, which makes it difficult to combine data taken at different epochs, radio data suggest the upstream edge of HST-1 may lie at the northern edge of the jet, possibly caused by an interaction between the jet and something in the external environment \citep{CHS07}.  We conclude that knot HST-1 is not the best choice for a pattern speed, and take the average radio-determined motion of knot D as the best available indicator of pattern speed.

To determine the evolution of pattern speed along the jet,  we  assume sufficiently slow spatial growth so that the filament wavelength, $\lambda_p$, can be related to the pattern speed by $\lambda_p \simeq 2 \pi v_p / \omega$. We assume that the wave frequency $\omega$ is set at the filament source, somewhere upstream of HST-1, and remains constant along the jet. The observed  50\% increase in $\lambda_p$ between HST-1 and knot A requires a 50\% increase in $v_p$  over this distance.   We tie this required increase to the average observed speed at knot D, $\beta^{\rm ob} = 0.40$. This choice requires ``observed'' pattern speeds of $\beta_{p}^{\rm ob} = 0.35$ at HST-1 and $\beta_{p}^{\rm ob} = 0.53$ at knot A, which are consistent with the ranges in the data, namely  $\beta^{\rm ob} \sim 0.2 - 0.8$, at HST-1 and  $\beta^{\rm ob} \sim 0.41 - 0.61$ at knot A \citep{BZO95}.

\vspace{-0.8cm} 
\subsubsection{Intrinsic pattern properties}
\vspace{-0.1cm}

\begin{figure}[h!]
\vspace{7.0cm}
\includegraphics{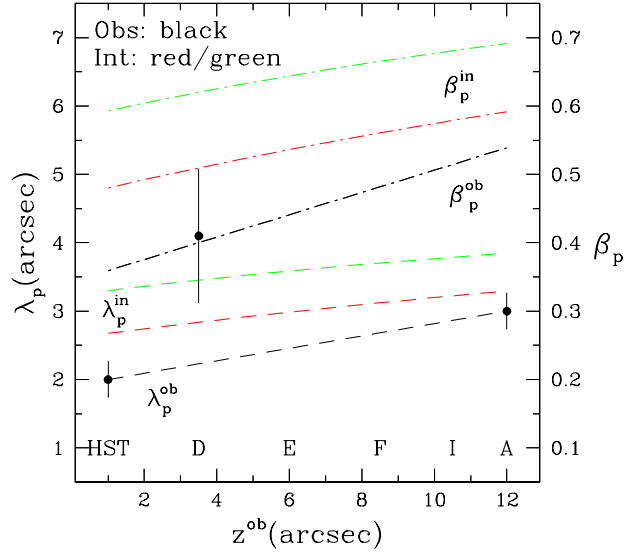}
\caption{\footnotesize \baselineskip 11pt Observed and intrinsic wavelengths for the inner M\,87 jet.  The observed wavelength $\lambda_p^{\rm ob}$ (black dashed line) and error bars are from LHE; we fit a linear relation between these two points. The observed pattern speed $\beta_p^{\rm ob}$  (black dash-dot line), and error bar, from \citep{BZO95}, refer to knot D. Note that the pattern speed increase is set by the observed wavelength increase; thus we fit to only one data point. The intrinsic wavelength,  $\lambda^{\rm in}_p$ (dashed lines), and pattern speeds, $\beta^{\rm in}_p$  (dash-dot lines), are derived for our {\it fast jet model} ($\theta = 15\arcdeg$; green) and our {\it slow jet model} ($\theta =   25\arcdeg$; red), given the observed wavelength, $\lambda^{\rm ob}_p$ and the observed the pattern speed, $\beta^{\rm ob}_p$.   Knot positions and nomenclature are indicated by the letters above  the lower axis.}
\label{Obs_vs_Int}
\end{figure}

Finally, we connect the  observed pattern wavelength, $\lambda^{\rm ob}_{p}$, to the intrinsic pattern wavelength, $\lambda^{\rm in}_{p}$, by 
\begin{equation}
 \lambda^{\rm ob}_{p} =  {\beta^{\rm ob}_p \over \beta^{\rm in}_p }
\lambda^{\rm in}_p ={ \sin\theta \over
1- \beta^{\rm in}_p \cos\theta}\lambda^{\rm in}_p~.
\end{equation}
Figure \ref{Obs_vs_Int} shows the intrinsic pattern speeds and wavelengths from HST-1 to knot A accompanying the observed wavelengths and observed pattern speeds  for the jet at 
$\theta = 15\arcdeg$ and at $\theta = 25\arcdeg$. In general, the observed wavelength can be greater or less than the intrinsic wavelength depending on the intrinsic pattern speed, $\beta^{\rm in}_p$, and viewing angle, $\theta$. 
For our viewing angles the intrinsic wavelength and pattern speed are greater than the observed values, and show less change along the jet  than the observed values.  In particular, the intrinsic wavelength,
$\lambda_p^{\rm in}$, increases more slowly than the jet radius, $r_j$, so that $\lambda^{\rm in}_p / r_j$ decreases along the jet. This has consequences for the KH instability analysis we present in Section \ref{KH_Modes}.

\vspace{-0.7cm}
\section{Models of the Jet Flow Field}
\label{flow_models}
\vspace{-0.1cm}

In this section we develop our baseline jet models, which we need in order to analyze the behavior of the KH instability, and determine what conditions in the jet would accompany deceleration in accord with the
observations.  We consider a steady-state jet  which conserves mass flux and is 
in pressure balance with an external cocoon. The cocoon acts as the ambient medium which controls the development of the KH instability.  The full system conserves energy flux as the jet decelerates. 

We do not tackle the complex problem of what causes the jet to decelerate.  Instead, we parameterize our 
jet models to describe the extent to which energy flux is conserved in the flow.   We begin with models in which energy flux is conserved within the jet, so that the jet plasma heats up as it decelerates.  We also explore jets in which a small fraction of the energy flux is transfered to the cocoon, as would be expected if the jet's interaction with the cocoon is responsible for the observed jet deceleration. Provided the energy lost from the jet is a small fraction of the total energy flux, both models give very similar results for the internal structure of the jet.

\vspace{-0.7cm}
\subsection{The Energy-Conserving Jet}
\label{EC_jet}
\vspace{-0.1cm}

We first consider a jet which conserves mass flux and energy flux. Because the jet expands uniformly between  HST-1 and knot A, we can assume flow along radial lines in spherical geometry. Conservation of mass flux can be written in terms of  two locations along the jet, a fiducial point at axial position $z_{j0}$
(with jet radius $r_{j0}$) and another point at some general $z_j$ (with jet radius $r_j$):
\begin{equation}
\gamma_j \rho_j \beta_j r_j^{2}=\gamma_{j0}\rho_{j0}\beta_{j0}r_{j0}^{2}~,
\label{MassCon}
\end{equation}
where $\rho_j$ is the jet mass density;   $\gamma_j$ and $\beta_j$ are the usual Lorentz
factor and normalized flow speed. Conservation of energy flux can be written as
\begin{equation}
\gamma_j^{2}W_j \beta_j r_j^{2}=\gamma_{j0}^{2}W_{j0}\beta_{j0}r_{j0}^{2}~,
\label{EnergyCon}
\end{equation}
where the total energy content of the plasma, the  ``relativistic" enthalpy (rest mass plus enthalpy), is described  by
\begin{equation}
W_j \equiv \rho_j +\left( {\Gamma_j  \over  \Gamma_j -1} \right) {P_j \over c^{2}}
\label{W_def}
\end{equation}
and $\Gamma_j$ is the adiabatic index of the jet plasma. We also define the ratio of pressure to rest mass energy, ${\Theta}_j = P_j / \rho_j c^2$, and the   specific enthalpy,
\begin{equation} 
\chi_j \equiv {\Gamma_j \over \Gamma_j-1} \frac{P_j}{\rho_j c^{2}} 
= { \Gamma_j \over \Gamma_j -1} {\Theta}_j ~.
\label{chi_def}
\end{equation}
We note that all of the jet parameters -- $\gamma_j(z)$, $\beta_j(z)$, $ P_j(z)$, $\rho_j(z)$, $\Gamma_j(z)$, $\chi_j(z)$ and ${\Theta}_j(z)$  -- are functions of position along the jet.  We suppress this $z$-dependence to lighten the notation.

Values for $\chi_j$ along an expanding jet containing relativistic plasma can vary from less than one to greater than one.  The adiabatic index, $\Gamma_j$, also changes with $\chi_j$ in this regime. We use the approximation \citep{T86,DHO96}
\begin{equation}
\Gamma \approx \frac{10.72+5\rho c^{2}/P}{8.04+3\rho c^{2}/P}~,
\label{G_approx}
\end{equation}
and use this to  connect  $\chi_j$ to ${\Theta}_j$ as
\begin{equation}
\chi_j ({\Theta}_j)
= {\Theta}_j f({\Theta}_j) 
\equiv {\Theta}_j  { \left( 10.72 {\Theta}_j + 5 \right) /  \left( 2.68 {\Theta}_j + 2 \right) } ~. 
\label{chi_R}
\end{equation}
The last equality defines the order-unity function $f({\Theta})$, which has the expected limits, $f({\Theta}) \to 2.5$ for ${\Theta} \ll 1$, and $f({\Theta}) \to 4 $ for ${\Theta} \gg 1$.

We now use Equation (\ref{MassCon}) in Equation (\ref{EnergyCon}) to rewrite energy flux conservation as
\begin{equation}
\gamma_j \left[ 1+\left({ \Gamma_j \over  \Gamma_j-1 } \right)
{P_j \over\rho_j c^{2}} \right] =\gamma_{j0}\left[ 1+\left({ \Gamma_{j0} 
\over \Gamma_{j0} -1 } \right) {P_{j0} \over \rho_{j0}c^{2}} \right]~,
\label{AllEnergyConserv}
\end{equation}
or 
\begin{equation}
\gamma_j \left[ 1+\chi_j({\Theta}_j) \right] =\gamma_{j0} \left( 1+\chi_{j0}\right)~, 
\label{EnergyCon2}
\end{equation}
where $\chi_{j0} \equiv \chi({\Theta}_{j0})$. This makes it apparent that a decelerating jet which conserves energy flux will heat up
($\gamma_j < \gamma_{j0} \Rightarrow \chi_j > \chi_{j0} \Rightarrow
{\Theta_j} > {\Theta}_{j0}$), and vice versa.

We want to solve Equation (\ref{EnergyCon2}) to describe the flow field in the jet.  In order to do this we need to connect the enthalpy, $\chi_j$, to the flow field.  Rather than modeling the full thermodynamics of the flow, we assume a polytropic dependence between the pressure and density,
\begin{equation}
{P_j \over P_{j0}} = \left( {\rho_j \over \rho_{j0}} \right)^{\varepsilon_{\rm ec}}
= \left( {\gamma_{j0} \beta_{j0} r_{j0}^2 \over \gamma_j \beta_j r_j^2} \right)^{\varepsilon_{\rm ec}} ~,
\label{basic_polytrope}
\end{equation}
for some (as yet unknown) index $\varepsilon_{\rm ec}$. The subscript ``ec'' denotes ``energy-conserving''. For an expanding jet, $\varepsilon_{\rm ec} = \Gamma_j >1 $ describes an accelerating adiabatic jet, $\varepsilon_{\rm ec} = 1$ describes an isothermal jet at constant speed, and $\varepsilon_{\rm ec} < 1$ describes a jet which heats up as it decelerates.  Our polytropic assumption also gives
\begin{equation}
{{\Theta}_j \over {\Theta}_{j0}} = \left( {\rho_j \over \rho_{j0}} \right)^{\varepsilon_{\rm ec} -1}
= \left( {\gamma_{j0} \beta_{j0} r_{j0}^2 \over \gamma_j \beta_j r_j^2} \right)^{\varepsilon_{\rm ec} -1} 
 \label{Theta_polytrope}
\end{equation}
and
\begin{equation}
{\chi_j ({\Theta}_j) \over \chi_{j0} } 
= \left( {\rho_j \over \rho_{j0}} \right)^{\varepsilon_{\rm ec} -1}g({\Theta_j})
= \left( {\gamma_{j0} \beta_{j0} r_{j0}^2 \over \gamma_j \beta_j r_j^2} \right)^{\varepsilon_{\rm ec} -1} 
g({\Theta_j}) ~. 
\label{chi1_polytrope}
\end{equation}
Here $g({\Theta_j}) = f({\Theta_j})/ f({\Theta}_{j0})$ is another order-unity function, which can be expressed in terms of $(\rho_j/\rho_{j0})$ and $\chi_{j0}$.

\vspace{-0.7cm}
\subsection{Constraints on Pressure Decline Along the Jet}
\label{pressure}
\vspace{-0.1cm}

Equations (\ref{Theta_polytrope}) and (\ref{chi1_polytrope}) show that the thermal state of the jet (described by $\chi$ or ${\Theta}$) is uniquely related to $\gamma$, if the polytropic index $\varepsilon_{\rm ec}$ is known. In particular, $\varepsilon_{\rm ec}$ determines the pressure drop along the jet (from Equation \ref{basic_polytrope}).  We therefore use the data to estimate $\varepsilon_{\rm ec}$.  Two sets of evidence are available: radio data which can be used to determine the minimum pressure in the jet and X-ray data which reveal the pressure in the local ISM.

Basic synchrotron theory can be used to find the minimum jet  pressure, $P_{\rm min}$, consistent with the jet's observed  luminosity and volume, e.g., \citet{P1970}; \citet{BOR1979}.  OHC  calculated $P_{\rm min}$ from their radio observations, ignoring  relativistic effects.  They found that $P_{\rm min}$ declines by a  factor $\sim 1.5$ between HST-1 and knot D, and by a further factor  $\sim 2$ between knots D and E. Because their data were obtained  when HST-1 was in a quiescent state, their $P_{\rm min}$ value for  that knot should be typical of the underlying jet.  The scatter in  their $P_{\rm min}$ estimates downstream of knot E suggests little  additional decline to knot A, and thus an overall decline in $P_{\rm  min}$ by a factor of $\sim 3$ between HST-1 and knot A.  Similar  calculations carried out by \citet{BSH91}, including optical and  X-ray data, obtained similar results.

\citet{SAK06} revisited $P_{\rm min}$ in the jet, including relativistic effects {\it via} a beaming correction $P_{\rm min} \propto \delta^{-10/7}$ \citep{SSO03}, where the Doppler factor is $\delta \equiv [\gamma( 1 - \beta \cos \theta)]^{-1}$.  In Section \ref{Jet_speed} we argued that typical jet Lorentz factors $\sim 4.4\!-\!7.5$ at HST-1, slowing to $\sim 3.1\!-\!5.4$ at knot D.  These values, along with their respective viewing angles 25\arcdeg~and
15\arcdeg, give Doppler factors $\delta \sim 1.9\!-\!3.1$ at HST-1 and $\delta \sim 2.2\!-\!3.6$ at knot D.  (The slightly larger values at knot D are a consequence of the $\beta \cos \theta$ dependence in $\delta$.) This result suggests an additional decline in $P_{\rm   min}$, relative to the uncorrected estimate, of $\sim 20 - 25\%$
from HST-1 to knot A.  Overall, these calculations suggest that $P_{\rm min}$ declines by a factor $\sim 3\!-\!4$ between HST-1 and knot A. There is, of course, no fundamental reason why the pressure of the jet plasma should be equal to $P_{\rm min}$.  In particular, if the jet is only weakly magnetized, we expect $P_j > P_{\rm min}$.  Nonetheless, we might guess that a decline in $P_{\rm min}$  approximately reflects the decline in $P_j$.  

The evolving jet and cocoon need not be in pressure balance with the ambient ISM and minimum pressure estimates (Section \ref{pressure_scaling}) indicate an overpressured jet. Even so, it is still interesting to consider the pressure drop in the ISM along the jet. On $\sim 10\!-\!100$ kpc scales, the ISM is fairly smoothly distributed, and its pressure declines smoothly with radius, e.g., \citet{NB1995}. However, the ISM in the inner part of the galaxy is complicated (as in Section \ref{flow_obs}), and the ISM pressure probably fluctuates substantially about any large-scale trend.   Nonetheless, \citet{SAK06} estimated a factor $\sim 3$ decline in the ISM pressure between HST-1 and knot A which is interestingly similar to the factor $\sim 3\!-\!4$ drop in $P_{\rm min}$ from the radio data.

Based on this evidence, we proceed by modeling a jet whose pressure declines by a factor $\sim 3-4$ between HST-1 and knot A. The value of $\varepsilon_{\rm ec}$ connects the pressure decline to the jet flow field. Using Equation (\ref{basic_polytrope}), we find $\varepsilon_{\rm ec} = 0.325 \pm 0.025$ provides a pressure decline $\sim3.5 \pm 0.5$ between HST-1 and knot A.  This result is nearly independent of the initial jet speed
and viewing angle used to fit the observed jet deceleration.  In what follows we set $\varepsilon_{\rm ec} = 0.325$ for our energy-conserving jet models. Note, we do not specify  an absolute value for the pressure at HST-1;  our KH analysis only needs the pressure change along the jet.  We discuss pressure scaling further
 in Section \ref{Scaling}.

\vspace{-0.7cm}
\subsection{Structure of the Energy-Conserving Jet}
\label{Jet_structure_models}
\vspace{-0.1cm}

Having chosen the polytropic index, we can build jet models which conserve energy flux and mass flux.  In Section \ref{EC_jet} we showed that the two parameters, $\chi_{j0}$ and $\varepsilon_{\rm ec}$, totally
specify the flow field in the jet, if the Lorentz factor, $\gamma_{j0}$, is known at some initial point $r_{j0}$. To solve for the flow field, we combine equations (\ref{EnergyCon2}) and (\ref{chi1_polytrope}), as
\begin{equation}
\gamma_{j0} ( 1 + \chi_{j0}) = \gamma_j \left[1 + \chi_{j0}
 \left( {\gamma_{j0} \beta_{j0} r_{j0}^2 \over \gamma_j \beta_j r_j^2} \right)^{\varepsilon_{\rm ec} -1} 
g({\Theta}_j) \right] ~.
\label{Eqn_to_Solve}
\end{equation}
This equation has only one unknown, $\gamma_j$, the jet Lorentz factor at radius  $r_j = \psi z_j$ (recall that $\psi$ is held constant and $g({\Theta_j})$ is an order-unity function which can be written in terms of the jet speed, radius, $\chi_{j0}$ and $\varepsilon_{\rm ec}$). In our solution, we begin at HST-1, choosing  $\varepsilon_{\rm ec} = 0.325$, and a specific set of initial conditions ($r_{j0}, \gamma_{j0}, \chi_{j0}$). We solve Equation (\ref{Eqn_to_Solve}) numerically, to determine $\gamma_j(z)$ at several positions along  the jet, out to knot A (where $r_j \sim 12 r_{j0}$).  We compare the results to the  values of $\beta^{\rm ob}_j$  at HST-1, knot D, knot E and the lower limit to $\beta^{\rm ob}_j$ at knot A, then adjust $\chi_{j0}$ to iterate on Equation (\ref{Eqn_to_Solve}) as needed, to give the best overall fit to the data for our chosen $\varepsilon_{\rm ec}$.  Finally,  we use Equations  (\ref{basic_polytrope}) or (\ref{Theta_polytrope}) to  find the evolution of the internal energy along the jet.

\begin{figure}[h!]
\vspace{6.5cm}
\includegraphics{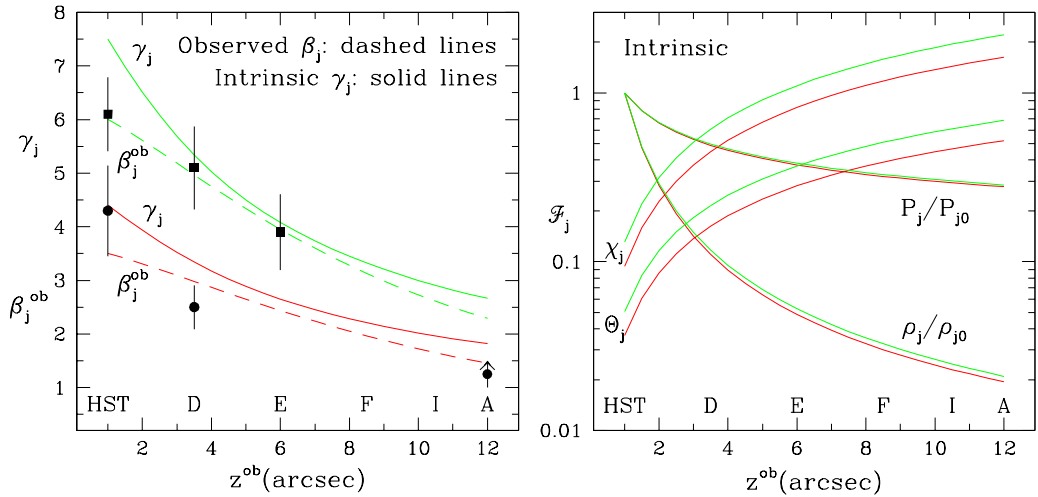}
\caption{\footnotesize \baselineskip 11pt Structure of the energy-conserving jet, for our {\it fast jet} (green lines) and {\it slow jet} (red  lines) models. Left: the observed jet speed speed,   $\beta^{\rm ob}_j$, (dashed lines) and intrinsic jet Lorentz factor,   $\gamma_j$, (solid lines). The optical speeds, $\beta^{\rm ob}$, at   HST-1, knot D and knot E are indicated by filled squares and radio speeds, $\beta^{\rm ob}$, at HST-1 and knot D are indicated by filled  circles, both with error bars.  The filled circle and arrow at knot A  indicate the lower limit to $\beta^{\rm ob}$ set by the fastest observed radio superluminal speed at knot B.  Right:  the specific enthalphy, $\chi_j$, normalized temperature $\Theta_j = P_j / \rho_j c^2$, pressure $P_j/P_{j0}$ and density $\rho_j/\rho_{j0}$, from solutions to Equation (\ref {Eqn_to_Solve}) with $\varepsilon_{\rm ec} = 0.325$.
\label{Jet_Flow}}
\end{figure}

In Figure \ref{Jet_Flow} we show our best fits to fast and slow energy-conserving jets. Our fast jet with $\theta=15\arcdeg$  starts at HST-1 with $\gamma_{j0} = 7.50$ and $\chi_{j0} = 0.130$.  By knot A it has
decelerated to $\gamma_{j} = 2.66$ and heated to $\chi_{j} = 2.189$. This model easily fits the optical superluminal speeds and significantly exceeds the lower limit at knot A.
 Our slow jet with $\theta=25\arcdeg$ starts at HST-1 with $\gamma_{j0} = 4.40$ and  $\chi_{j0} = 0.095$.  By knot A it has decelerated to $\gamma_{j} = 1.82$ and heated to  $\chi_{j}  = 1.643$.  This model is just barely  acceptable; it lies just outside the error bars associated with the fastest radio superluminal speeds at HST-1 and knot D, and just above the lower limit at knot A. We therefore treat our slow-jet model as a lower limit to the jet speed and upper limit to the viewing angle. In both fast and slow models the enthalpy rises from $\sim 0.1$ at HST-1 to $\sim 2.0$ at knot A; the jet plasma heats as it absorbs the lost kinetic energy.  The jet plasma is hot but subrelativistic, $P \sim 0.04 \rho c^2$, at HST-1 and heats by a factor $\sim 15$ to $P \sim 0.6 \rho c^2$ at knot A.  We do not derive numerical values for the pressure or density, but only the ratios of each quantity to its value at HST-1; these are also shown in Figure \ref{Jet_Flow}. Although the jet decelerates, its density still drops due to the expansion, by a factor $\sim 50$ between HST-1 and knot A.  The jet pressure is, of course, constrained by our assumptions to drop by a factor $\sim 3.5$ over the same range.

Solutions to Equation (\ref{Eqn_to_Solve}) are sensitive to the starting value
of the enthalpy, $\chi_{j0}$.  Flows in which the jet plasma is internally relativistic at HST-1 ($\chi_{j0} \gg 1$) decelerate very rapidly and violate the lower limit on jet speed at knot A.  Flows which are very cool at HST-1 ($\chi_{j0} \ll 1$) decelerate too slowly to match  the proper motions through knots D and E. Specifically, we find an upper limit to the starting enthalpy, $\chi^{\rm max}_{j0} = 0.74$ by considering a maximum allowed deceleration from $\gamma_{j0} = 10.5$ at HST-1 to $\gamma_{j} = 1.8$ at knot A (using the upper error bar limit, $\beta^{\rm ob} \le 6.7$ at HST-1 and the lower error bar limit $\beta^{\rm ob} \ge 1.12$ at knot A). We also find a lower limit of $\chi^{\rm min}_{j0} = 0.032$, by considering a minimum allowed deceleration from $\gamma_{j0} = 6.3$ at HST-1 to $\gamma_j = 4.7$ at knot E (using the lower error bar limit, $\beta^{\rm ob} \ge 5.5$, at HST-1 and the upper error bar limit, $\beta^{\rm ob} \le  4.5$, at knot E). The minimum deceleration case gives $\gamma_j = 3.3$ at knot A. 

\vspace{-0.7cm}
\subsection{The Energy-Losing Jet}
\label{EL_jet}
\vspace{-0.1cm}

In our energy-conserving jet, all of the excess kinetic energy lost as the jet decelerates  is converted into internal energy within the jet. Because the jet probably decelerates through interactions with the surrounding medium, some of that excess kinetic energy may well go into the surroundings.  This could happen, for instance, as a consequence of shocks driven into the cocoon by the unstable KH modes.  We want to explore how such energy loss affects  the  KH instability.  However, building a complete model of jet and ISM energetics is beyond the scope of this paper.  Instead, we invent a toy model to describe the effect of the energy transferred  from the jet to the cocoon.

We know that the pressure in an energy-losing jet must drop more rapidly, relative to the local density, than in an energy-conserving jet.  However, the ``observed'' jet pressure must still drop by a factor $\sim 3.5$, to match our arguments in Section \ref{pressure}.  We parameterize this as follows. We first assume that the underlying jet flow satisfies equations (\ref{MassCon}) and (\ref{EnergyCon}), and the pressure also obeys a polytropic relation, but now with an ``energy-losing'' polytropic index, $\varepsilon_{\rm el} < \varepsilon_{\rm ec}= 0.325$.
We solve Equation (\ref{Eqn_to_Solve}) with this new $\varepsilon_{\rm el}$, and match the solution to the observed velocities. We adopt the same Lorentz factors at HST-1 (7.50, fast jet, and 4.40, slow jet), and at knot A (2.66, fast jet and 1.82, slow jet), as for an energy-conserving jet, and we again adjust $\chi_{j0}$ to match these data. The values of $\varepsilon_{\rm el}$ and $\chi_{j0}$ consistent with the observed deceleration are interrelated;  smaller values of $\chi_{j0}$ accompany smaller values of $\varepsilon_{\rm el}$.  At this point, to ensure that the ``true'' jet pressure drops more rapidly than $\varepsilon_{\rm el}$ describes, we assume  the ``true'' jet pressure is related to the jet density {\it via} Equation (\ref{basic_polytrope}), but now with the 
energy-conserving $\varepsilon_{\rm ec} = 0.325$.  Thus, we parameterize the mismatch between the ``observed'' pressure decline, and an energy-conserving pressure decline, by $\varepsilon_{\rm ec} - \varepsilon_{\rm el}$. We return to these models in  Section \ref{limiting_jets}, where we will show that  energy-losing models consistent with the data must have $\varepsilon_{\rm el} > 0.250$, and that models with $0.250 < \varepsilon_{\rm el} < 0.325$ differ only  slightly from the energy-conserving models. 

\vspace{-0.7cm}
\section{KH Instabilities in the Jet-Cocoon System}
\label{KH_Modes}
\vspace{-0.1cm}

In this section we determine under what conditions the twisted filaments seen in the M\,87 jet can be caused by the KH instability. We begin with a brief overview of the KH instability, including some useful analytic approximations; we store more details in the Appendix. We analyse our energy-conserving and energy-losing jet
models, finding what conditions must exist in the cocoon if the KH-generated filaments in our model jets are to match the ones in the real jet.  We show that the KH modes propagate supersonically in the jet, and verify that they grow rapidly enough to saturate close to the origin of the jet.

KH  stability analysis describes the growth of perturbations to the jet  using solutions of the dispersion relation, Equation (\ref{DispRel})  in the Appendix.  This equation can be solved to determine the  wavelength, $\lambda_w$, and wave speed, $v_w$, associated with  a frequency, $\omega_w$.  The dispersion relation solution  describes a 180\arcdeg~rotation of the elliptical cross section  distortion, corresponding to a 180\arcdeg~rotation of a single  filament. Thus, the elliptical wave frequency  and wavelength described in solutions to the dispersion relation are related to the observed frequency and wavelength  by $\omega_w = 2 \omega_p$ and $\lambda_w = \lambda_p/2$.

Numerically, we rely on the intrinsic pattern   wavelength, $\lambda_p$, and speed, $v_p$, derived from the data (as  in Section \ref{flow_obs}), to determine the associated pattern  frequency, $\omega_p = 2 \pi v_p / \lambda_p$.  Specifically, from  Figure \ref{Obs_vs_Int}, the intrinsic pattern speeds at HST-1 are  $v_p \sim 0.60c$ for the fast jet, and $v_p \sim 0.48c$ for the slow  jet; the intrinsic wavelengths are $\lambda_p \sim 3.3\arcsec$ for  the fast jet and $\sim 2.7\arcsec$ for the slow jet. From these we  estimate $\omega_p \sim 1.4 \times 10^{-10}$~rad s$^{-1}$,  corresponding to a periodicity of $\sim  1440$~yr.  We assume $\omega_p$ is set at the filament source, somewhere  upstream of HST-1, and remains constant along the jet. 

We assume the jet is in pressure balance with the cocoon, as required by our KH analysis.  Solutions to the dispersion relation (see details in the Appendix) depend on the flow speed and thermal state of the jet
and the cocoon.  In practice, at a given location along the jet, we use the jet parameters from our models in Section \ref{flow_models} and iterate on the cocoon parameters until we find a good solution.
Specifically, we begin with an initial guess for the cocoon  sound speed, solve Equation (\ref{DispRel}) numerically for the  complex wavenumber and compute the associated wavelength  $\lambda(\omega)$ and wave speed, $v(\omega)$ as a function of  $\omega$ over a broad frequency range.  We iterate on $a_c$ until we
  obtain a value of $v_w = v_p$ at the wavelength $\lambda_w =  \lambda_p/2$ that is within 2\% of the observed values. This  typically requires five to ten iterations. Note that we are matching  computed wavelengths and wave speeds to observed intrinsic  wavelengths and wave speeds, and the frequency will be $\omega_w =
  \omega_p/2$.  We do this at seven locations along the jet: HST-1, 2\arcsec, knot D, knot E, knot F, knot I and knot A.

\vspace{-0.7cm}
\subsection{Analytic Approximations Close to Resonance}
\label{DR_analytic}
\vspace{-0.1cm}

KH instabilites are characterized by a resonant frequency and wavelength at which the growth rate is a maximum.  In general we expect perturbations to the jet to produce structures close to this resonance. In Section
  \ref{limiting_jets} we find that the elliptical surface mode  in our model jets is close to resonance.  Had we found the alternative -- that the observed filaments were   not close to resonance --  our identification of the  filaments with KH instability would be called into question.

While the KH dispersion relation must be solved numerically, as we do in Sections \ref{limiting_jets} and \ref{intermediate_jet}, analytic expressions for the resonant conditions provide insight into the numerical
results. The resonant solutions, given in Equations (\ref{resonant_wavespeed}) and (\ref{resonant_freq}) in the Appendix, depend on conditions in the jet and the cocoon. These expressions can be simplified for conditions found in our numerical solutions, namely,  $u_c \ll u_j \sim c$.  In this limit the resonant wave (pattern) speed becomes
\begin{equation}
{v_w^{\ast} \over u_j } \simeq 
{\gamma_j a_c \over \gamma_j a_c + a_j}
\label{approximate_wavespeed}
\end{equation}
Here, $u_j = \beta_j c$ is the jet speed, $\gamma_j^2 = ( 1 - \beta_j^2)^{-1}$, and $a_j$ and $a_c$ are the sound speeds in the jet and cocoon.  The sound speed depends on the thermal state of the plasma:
\begin{equation}
\beta_{a}^2 = { a^2\over c^2} = (\Gamma -1) { \chi \over (1 + \chi)}
= { \Gamma \Theta \over 1 + \Gamma \Theta / (\Gamma -1)} ~.
\label{sound_speed}
\end{equation}
In Equation (\ref{approximate_wavespeed}) we have also used the fact that $\gamma_a \simeq 1$ (because $a < 0.5c$).  From Equation (\ref{approximate_wavespeed}) we verify that resonant waves  move more slowly than the underlying flow:  $v_w^{\ast} < u_j$. We also see  that a change in the intrinsic wave speed, $v_{w}^*$, requires a change in the  cocoon sound speed, $a_c$, which  depends on the jet model  through $\gamma_j$ and $a_j$.  The resonant wave frequency becomes
\begin{equation}
{ \omega_{w}^{\ast} r_j \over a_c} \simeq { 5 \pi \over 4} 
\left[ 1 - \left( { a_c \over v_w^{\ast}} \right)^2 \right]^{-1/2}
\label{approximate_res_freq}
\end{equation}
where $r_j$ is the local jet radius. We can combine Equations(\ref{approximate_wavespeed}) and
(\ref{approximate_res_freq}) with $\lambda^{\ast}_w = 2 \pi v_w^{\ast} / \omega^{\ast}$ to get
\begin{equation}
{ \lambda^{\ast} \over r_j} \simeq { 8 \over 5} { u_j \over a_c} 
{ \left[ 1 - (a_c/v_w^{\ast})^2 \right]^{1/2}
\over \left[ 1 + ( a_j / \gamma_j  a_c) \right]}
\label{approximate_lambda}
\end{equation}
Equation (\ref{approximate_lambda}) shows that the cocoon thermal state -- described here by the sound speed -- is the {\it only} important parameter (given $u_c \ll a_c$).  If $\lambda^*/r_j$ decreases,  $u_j / a_c$ must decrease.  But, because $u_j \sim c$,  $a_c$ must rise in order to
account for the filaments in terms of a KH instability. 

\vspace{-0.7cm}
\subsection{Limiting-Case Jet and Cocoon Models} 
\label{limiting_jets}
\vspace{-0.1cm}

In addition to energy-conserving jets (Section \ref{Jet_structure_models}), we want to analyze the energy-losing jets (Section \ref{EL_jet}).  In the latter case the cocoon as well as the jet is involved in conservation of energy flux, so we replace Equation (\ref{EnergyCon}) by
\begin{equation}
\gamma _{c}^{2}W_{c}\beta _{c}(r_{c}^{2} - r_j^2) + \gamma_j^{2}W_{j}\beta_jr_j^{2}
=  \gamma_{j0}^{2}W_{j0}\beta_{j0}r_{j0}^{2} +\gamma _{c0}^{2}W_{c0}\beta_{c0}
(r_{c0}^{2} - r_{j0}^2)
\label{two_phase_energy}
\end{equation}
where the subscript ``$c$'' labels properties in the  cocoon. For the energy-losing jet models, we self-consistently compute $\beta_c$ from Equation (\ref{two_phase_energy}), as part of the iteration to solve 
for $a_c$.  We assume that the observed jet carries essentially all the energy flux at HST-1: $\gamma_{c0}^{2}W_{c0}\beta _{c0}(r_{c0}^{2} - r_{j0}^2) \ll \gamma_{j0}^{2}W_{j0}\beta_{j0}r_{j0}^{2}$.  We find that the cocoon flow speed is dynamically insignificant, $\beta_c \ll \beta_{ac}$, provided  the cocoon radius $r_c > 4 r_j$;  in specific models  we assume $r_c = 10 r_j$.

We can only find physical solutions to cocoon conditions in the energy-losing jets if  $0.250 \le \varepsilon_{\rm el} \le 0.325$.   At smaller values of $\varepsilon_{\rm el}$ the cocoon sound speed necessary to match the filament properties goes to zero somewhere between HST-1 and knot E, requiring the cocoon density to become infinite in order to maintain pressure balance. The lower limit on $\varepsilon_{\rm el}$ corresponds to an upper limit on energy lost from the jet to the cocoon, $\sim 19$\% for fast jets and $\sim 16$\% for slow jets. In this section we take $\varepsilon_{\rm el} = 0.250$ for our energy-losing jets, to determine the effects of the largest possible energy loss.

\begin{figure}[h!]
\vspace{12.5cm}
\includegraphics{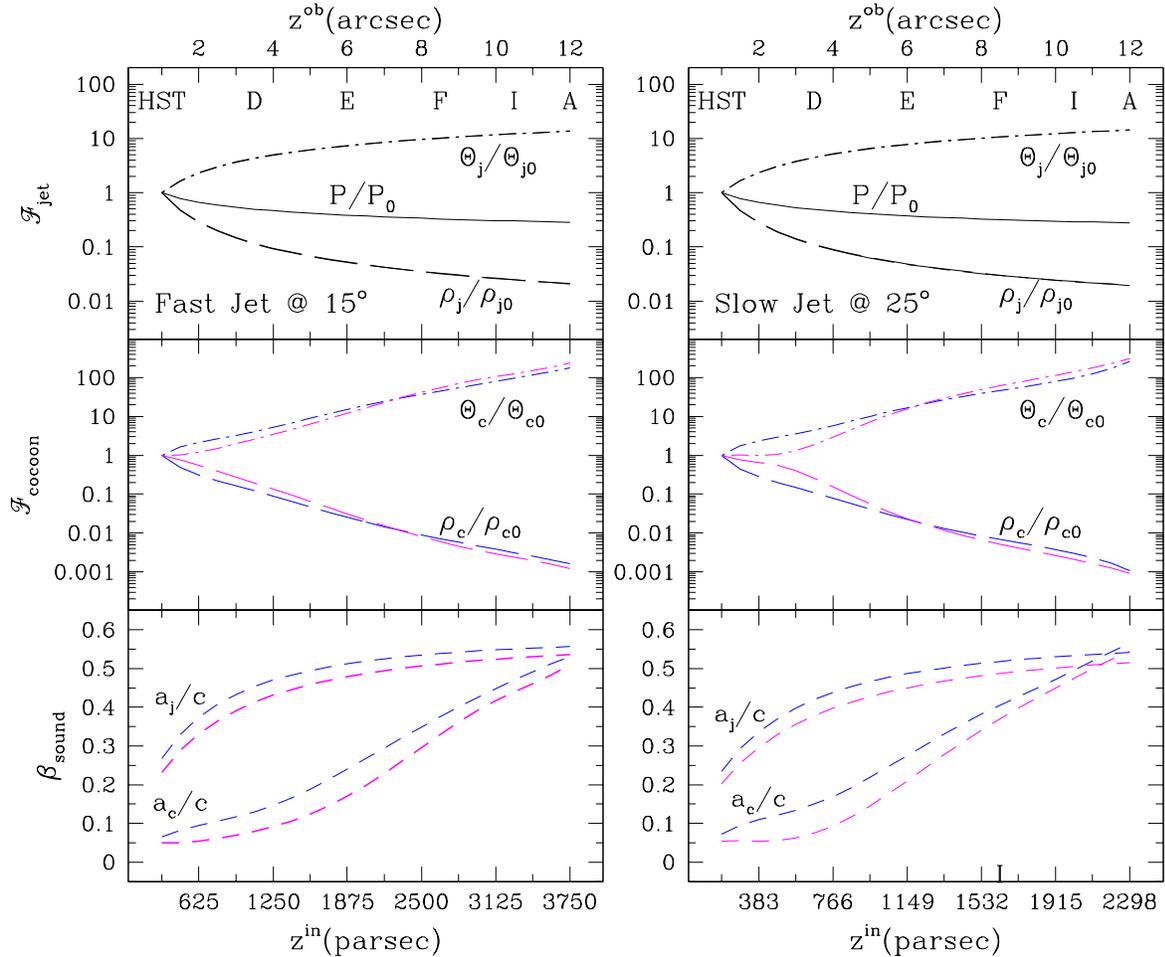}
\caption{\footnotesize \baselineskip 11pt Jet and cocoon parameters for the fast jet models (left) and slow
jet models (right), as a function of intrinsic distance along the  jet. Energy-conserving jets are shown in blue; maximum energy-losing jets in magenta. The upper panels show the jet  temperature,  relative pressure and relative number density.  The  middle panels show the cocoon temperature and relative number  density. The bottom panels shows sound speeds in the jet, $a_j/c$, and cocoon, $a_c/c$.}
\label{Jet_Models}
\end{figure}

We show the results of our KH analysis in Table 1 and in Figure \ref{Jet_Models}. Table 1 gives values 
at the fiducial point, HST-1, for our four limiting-case jet models (fast and slow, energy-conserving and energy-losing). The thermal state of the jet is described {\it via} the enthalpy, $\chi_{j0}$, the adiabatic index, $\Gamma_{j0}$, and the sound speed $\beta_{aj,0}$.  The thermal state of the cocoon is given {\it via} the same three quantities. 
\begin{table}[h!]
\vspace{-0.25cm}
\caption{Jet, Cocoon \& Elliptical Mode Parameters at HST-1}
\begin{center}
\begin{tabular}{lcccccccccc}
\tableline
\noalign{\smallskip}
Model & $z$(pc)&$\chi_{j0}$&$\Gamma_{j0}$&$\beta_{aj,0}$&$\chi_{c0}$&$\Gamma_{
c0}$&$\beta_{ac,0}$&$\omega_w r_j/u_j$&$\lambda_w/r_j$&$\lambda_{w}^*/r_j$ \\
\noalign{\smallskip}
\tableline
F$_{\rm ec}$& 312 & 0.130 & 1.627 & 0.268 & 0.0064 & 1.664 & 0.065 & 0.119 & 31.5 & 31.5 \\
F$_{\rm el}$& 312 & 0.092 & 1.637 & 0.232 & 0.0038 & 1.665 & 0.050 & 0.119 & 31.5 & 38.6 \\
\tableline
S$_{\rm ec}$& 192 & 0.095 & 1.637 & 0.235 & 0.0079 & 1.664 & 0.072 & 0.122 & 25.5 & 23.4 \\
S$_{\rm el}$& 192 & 0.068 & 1.644 & 0.203 & 0.0046 & 1.665 & 0.055 & 0.122 & 25.5 & 28.5 \\
\tableline
\end{tabular}
\end{center}
\vspace{-0.4cm}
\end{table}
Table 1 also contains the wave frequency, $\omega_w$, and wavelength, $\lambda_w$, that fits the observed intrinsic wavelength and wave speed along with the resonant wavelength, $\lambda_{w}^*$, of the elliptical KH mode computed from the dynamical and thermal state of the jet and cocoon.  Quantities are scaled to the jet radius, $r_j$, and speed, $u_j$, at HST-1. We see that the intrinsic filament wavelengths at HST-1, if produced by the elliptical surface mode, are within 20\% of the resonant wavelength for that mode.   This is in accord with our expectation that perturbations will  produce structures close to the resonant wavelength.

In the upper panels of Figure \ref{Jet_Models} we show the density, pressure and internal energy for  both fast and slow versions of both jet models, interpolated with a seventh-order polynomial between the numerical solutions at the seven locations along the jet.  The middle panels of Figure \ref{Jet_Models} show the cocoon density and temperature structure required to explain the filaments in the jet in terms of  the KH instability.  Because we assume pressure balance, the pressure decline in the upper panels of Figure \ref{Jet_Models} also applies to the cocoon. The lower panels of Figure \ref{Jet_Models} highlight the internal energy in the jet and cocoon,  {\it via} the sound speed in each plasma. In both fast and slow models, the cocoon starts cooler than the jet, but heats up more rapidly, so that $a_c \sim a_j$ by knot A. 

One key result is the thermal state of the cocoon required in order for the observed filaments to be caused by the KH instability.  At HST-1, the cocoon must be cooler than the jet, but still quite warm; $\chi_{c0} \sim 0.004 - 0.008$, or $P_{c0} / \rho_{c0} c^2 \sim 0.002 - 0.003$.  Because the jet and cocoon are in pressure balance, the
cocoon must be denser than the jet at HST-1: $\rho_{c0} / \rho_{j0} \simeq \chi_{j0} /\chi_{c0} \sim 12-24$.  By knot A the internal energy of the cocoon must be nearly relativistic, and comparable that of the jet: $\chi_{cA} \sim 1.3 - 2.4$, or $P_{cA}/\rho_{cA}c^2 \sim 0.4 - 0.7$. Pressure balance requires $\rho_{cA} \sim \rho_{jA}$ at
knot A. 

We emphasize that the increase in cocoon temperature along the jet is not a result of energy deposition from the decelerating jet.  Rather, the temperature increase is required to explain the observed filament wavelength, which increases less rapidly than the jet radius (e.g., Figure \ref{Obs_vs_Int}).  As discussed in Section \ref{DR_analytic}, the decrease of $\lambda_w / r_j$ along the flow requires $u_j/a_c$ to decrease as well, and thus $a_c$ must increase because $u_j \simeq c$ throughout the jet.  We do not try to
  model the thermal state of the cocoon; such modeling is beyond the   scope of this paper.  We simply note that such a hot cocoon must   exist if the observed filaments are caused by the KH instability.
  
\vspace{-0.7cm}
\subsection{Intermediate-Case Jet and Cocoon Models}
\label{intermediate_jet}
\vspace{-0.1cm}

Figure \ref{Jet_Models} shows  only small differences between the two limiting cases of energy-conserving and energy-losing jets. To continue, therefore, we  develop an intermediate case, $\varepsilon_{\rm int} = 0.300$, corresponding to an energy loss of 7\% for the fast jet and 6\% for the slow jet.  We develop fast and slow 
flow models for this intermediate-case jet, as in Section \ref{EL_jet}, and use these models for specific calculations in the rest of the paper. We carry out the filament-matching KH analysis, as  in Section \ref{limiting_jets};  we give the details of these  models  in Table 2.  
\begin{table}[h!]
\vspace{-0.2cm}
\caption{Jet, Cocoon and Elliptical Mode Parameters for the Intermediate-Case Jet} 
\vspace{-0.4cm}
\begin{center}
\begin{tabular}{lccccccccccc}
\tableline
\noalign{\smallskip}
Model &$z$(pc)&$\beta_j$& $\chi_j$&$\rho_j/\rho_{j0}$&$\beta_{aj}$&$\chi_c$&$\rho_c/\rho_{c0}$
&$\beta_{ac}$&$\omega_wr_j/u_j$&$\lambda_w/r_j$&$\lambda_{w}^*/r_j$ \\
\noalign{\smallskip}
\tableline
F$_{@HST}$ &312 & 0.991 & 0.115 & 1.000 & 0.255 &0.006& 1.000 & 0.061 & 0.119 & 31.5 &33.5 \\
F$_{@2\arcsec}$&625 & 0.988 & 0.285 & 0.289 & 0.359 &0.010& 0.386 & 0.080 & 0.239 & 16.0 &22.2 \\
F$_{@kntD}$ &1090& 0.982 & 0.567 & 0.120 & 0.439 &0.019& 0.150 & 0.110 & 0.420 & 9.41 &14.8 \\
F$_{@kntE}$ &1880& 0.970 & 1.048 & 0.052 & 0.501 &0.078&0.028 & 0.215 & 0.730 & 5.70 &8.19 \\
F$_{@kntF}$ &2660& 0.954 & 1.517 & 0.033 & 0.530 &0.279& 0.007 & 0.360 & 1.05 & 4.16 &4.38 \\
F$_{@kntI}$ &3280& 0.939 & 1.878 & 0.025 & 0.543 &0.642& 0.003 & 0.460 & 1.32 & 3.44 &2.48 \\
F$_{@kntA}$ &3750& 0.927 & 2.140 & 0.021 & 0.550 &1.291& 0.002 & 0.525 & 1.53 & 3.06 & ----- \\
\tableline
S$_{@HST}$ &192& 0.974& 0.085 & 1.000 & 0.224 &0.007& 1.000 & 0.068 & 0.122 & 25.5 &24.9 \\
S$_{@2\arcsec}$&383& 0.967& 0.206 & 0.282 & 0.323 &0.012& 0.371 & 0.090 &0.246 & 13.1 &16.0 \\
S$_{@kntD}$ &670& 0.954& 0.405 & 0.110 & 0.406 &0.024& 0.138 & 0.125 & 0.436 & 7.73 &10.5 \\
S$_{@kntE}$ &1150& 0.926& 0.737 & 0.049 & 0.475 &0.120& 0.022 & 0.260 & 0.770 & 4.75 &5.23 \\
S$_{@kntF}$ &1630& 0.892& 1.052 & 0.030 & 0.509 &0.355& 0.007 & 0.390 & 1.13 & 3.50 &1.83 \\
S$_{@kntI}$ &2010& 0.861& 1.288 & 0.023 & 0.525 &0.858& 0.003 & 0.490 & 1.45 & 2.93 &----- \\
S$_{@kntA}$ &2300& 0.836& 1.454 & 0.020 & 0.534 &2.381& 0.001 & 0.560 & 1.71 & 2.62 &----- \\
\tableline
\end{tabular}
\end{center}
\vspace{-0.3cm}
\end{table}
We again find the filaments are close to resonance over most of the jet.
However, Table 2 does not give a resonant wavelength at  knot A for either case, or at knot I for the slow jet case. Stability analysis \citep{G65} shows that the resonance which leads to a  maximum in the growth rate does not exist below some supersonic limit which depends on the jet Lorentz factor and sound speeds. For example, Equations (\ref{resonant_freq}) or (\ref{approximate_res_freq}) show a  rapid rise in the resonant frequency  as $a_c \rightarrow v^*_w < u_j$ (noting that $\gamma_j (\gamma_{a,c}a_c) \sim (\gamma_{a,j} a_j )$). Because $\omega^* \to \infty$ in this situation, the growth rate continues to increase in accord with the low-frequency approximation, Equation (\ref{low_freq_DR}), at all frequencies.

We can use this intermediate-case jet to show that the elliptical mode speed is supersonic in both cocoon and jet at HST-1 and is weakly supersonic in both the cocoon and the jet at knot I.  That is,  the jet plasma moves supersonically through the high pressure filaments associated with the elliptical mode, and the filaments themselves move supersonically relative to the cocoon plasma.
We demonstrate this in Table 3 for fast and slow versions of the intermediate jet where we give the relativistic 
\begin{table}[h!]
\vspace{-0.3cm}
\caption{Wave Speeds and Mach Numbers for the Intermediate-Case Jet}
\begin{center}
\begin{tabular}{lccccccc}
\tableline
\noalign{\smallskip}
Model &$\beta_j$&$\beta_{w,c}$&$\beta_{j,w}$&$\beta_{a,j}$&$\beta_{a,c}$&$M_{j,w}$&$M_{w,c}$ \\
\noalign{\smallskip}
\tableline
F$_{@HST}$  &  0.991   &   0.593 & 0.965 & 0.255  &   0.061  & 14.0  &   12.1 \\
F$_{@D}$      &   0.982   &   0.620 & 0.925 & 0.439  &   0.110  &   5.0  &   7.2 \\
F$_{@E} $      &  0.970   &   0.644 & 0.869 & 0.501  &   0.215   &  3.0   &   3.8 \\
F$_{@F} $      &  0.954   &   0.665 & 0.791 & 0.530   &  0.360   &  2.1   &   2.3 \\
F$_{@I} $       &  0.939   &   0.681 &  0.716 & 0.543   &  0.543   &  1.6   &   1.8 \\
\tableline
S$_{@HST}$  &   0.974   &   0.480&  0.928 &   0.224      &   0.068      &   10.8      &   9.0 \\
S$_{@D}$     &   0.954  &   0.509   & 0.865&   0.406       &   0.125       &   3.9      &   4.7 \\
S$_{@E}$     &   0.926   &   0.536   & 0.774 & 0.475       &   0.260       &   2.3     &   2.3 \\
S$_{@F} $    &   0.892   &   0.561   &  0.663 &0.509       &   0.390       &   1.5       &   1.6 \\
S$_{@I} $    &   0.861   &   0.579   &  0.562 &0.525       &   0.490       &   1.1       &   1.3 \\
\tableline
\end{tabular}
\end{center}
\vspace{-0.4cm}
\end{table}
Mach numbers for the wave speed relative to the cocoon, $M_{w,c}$ $\equiv (\gamma_{w,c}\beta_{w,c})/(\gamma_{a,c} \beta_{a,c})$, and for jet flow through the wave in the jet reference frame, $M_{j,w}$ $\equiv (\gamma_{j,w} \beta_{j,w})/(\gamma_{a,j} \beta_{a,j})$. Thus, the KH wave mode can produce supersonic shocks in both the cocoon and the jet at HST-1, and weakly supersonic shocks at knot I or knot F. This supports our conjecture that the KH instability in the jet decelerates and heats the jet, and dissipates some energy in the cocoon, by means of shocks generated by the filaments as they reach large amplitudes.
 
\vspace{-0.7cm}
\subsection{Normal Mode Growth Rates}
\label{growth_rates}
\vspace{-0.1cm}

In the previous sections we determined conditions that must exist in the cocoon if the elliptical surface mode of the KH instability is responsible for the filaments in the M\,87 jet.  In this section we verify that the growth rate of this mode is sufficiently rapid to saturate along the jet.  We also investigate whether other modes, in harmonic
resonance with the elliptical mode, might have comparable growth rates.  We work with the two intermediate-case jet models from Section \ref{intermediate_jet}.  We use the full dispersion relation to compute the spatial growth length, $\ell(z) = |k_i(z)|^{-1}$, where $k_i$, the spatial growth rate, is the imaginary part of the wavenumber associated with the elliptical mode frequency, $\omega_w = 2 \pi v_w /\lambda_w$.  We use continuously variable conditions in the jet and cocoon found by interpolation between the seven locations shown in Table 2. We also calculate the spatial growth rates of the helical, triangular and rectangular modes, at the harmonically resonant frequencies $\omega_H = \omega_w/2$, $\omega_E = \omega_w$, $\omega_T = 3\omega_w/2$ and $\omega_R = 2 \omega_w$. These harmonic frequencies produce  filaments with wavelength and wave speed comparable, but not identical, to the elliptical mode. We chose these frequencies based on numerical simulations which show that other normal modes are excited in frequency resonance with the dominant mode \citep{HCR97}, which we assume is the elliptical surface mode. Because we have found that the elliptical surface mode is operating near to resonance along the jet, the assumption that other surface modes operate at the specified frequency resonances  guarantees that these surface modes are  close to their fastest growing frequencies along the jet, i.e., $\omega^*_n \sim n \omega^*_H$ where $n = 1,2, ...$ is the helical, elliptical, etc.\ mode number.

\begin{figure}[h!]
\vspace{7.4cm}
\includegraphics{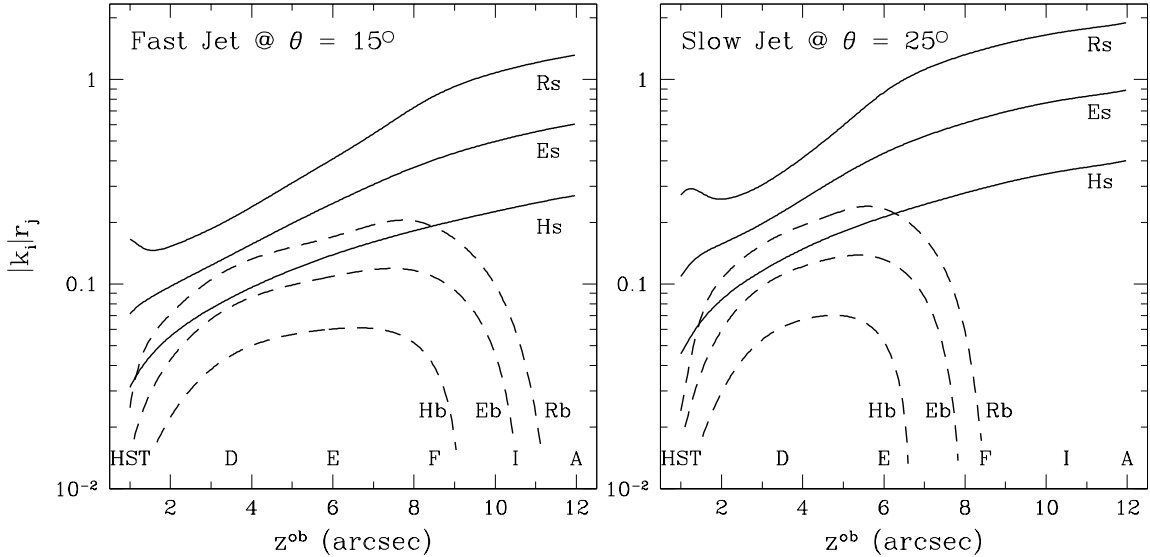}
\caption{\footnotesize \baselineskip 11pt Spatial growth rates, $|k_i| r_j$, of harmonically resonant rectangular (R), elliptical (E) and helical (H) modes, as a function of distance along the jet. Surface (s) modes are shown by solid lines;  body (b) modes are shown by dashed lines. The left panel shows the fast jet and the right panel shows the slow jet, both for the intermediate-jet model. } 
\label{Spatial_growth}
\end{figure}

In Figure \ref{Spatial_growth} we plot the spatial growth rate, $|k_i|$ normalized by $r_j = \psi z_j$, as a function of position along the jet.  We find  that each surface mode grows more rapidly than the accompanying body mode along the jet. The spatial growth rate of the surface modes when scaled to $r_{j0}$ at HST-1 decreases slightly along the jet. The triangular mode growth rate is not shown in this figure, for reasons of clarity; it lies  between the rectangular 
and elliptical mode growth rate. All four modes, however, are  used in the pseudo-synchrotron image calculated in Section \ref{Pseudo_Synch}.  Figure \ref{Spatial_growth} also shows that  surface modes will develop rapidly along the M\,87 jet;  growth lengths for the elliptical surface mode range from $\ell \gtrsim 10 r_{j0}$ at HST-1 to $\ell \gtrsim r_j = 12 r_{j0}$ at knot A. We therefore speculate that surface modes can grow to significant amplitudes. 

Although we find that higher order surface modes initially grow more rapidly than the elliptical mode, we argue that the elliptical mode dominates in the actual M\,87 jet.  The real jet must be bounded by an outer velocity shear layer.  Linear analysis shows that higher-order modes can be significantly affected, or even stabilized, by the
presence of a shear layer \citep{FMT82}, but the helical and elliptical surface modes remain unstable even in the presence of significant shear \citep{B91}.  In addition, simulations show that higher-order modes initially grow rapidly, but then saturate and are overwhelmed by more slowly growing, lower-order normal modes that
saturate at a larger amplitude \citep{HCR97}.  The spatial  growth rate of the helical surface mode is typically half that of   the elliptical surface mode.  This is    enough for the helical mode to overwhelm a saturated elliptical mode, if it is excited at all.  Thus, any initial helical perturbation to the M\,87 jet must have been very tiny.

We find that frequency-resonant body modes are also unstable in the inner region of the jet (Figure \ref{Spatial_growth}) but they grow  more slowly than surface modes.  The very low growth rates
of the body modes around HST-1, and also in the outer part of the jet, occur because the body modes are near their minimum unstable frequency in these regions. Complete stabilization of the body modes at all
frequencies occurs between knots I and A for the fast jet model and between knots F and I for the slow jet model, because the jet is not sufficiently supersonic in those regions \citep{H07}.  Based on this we argue  that body modes do not develop to significant amplitudes on the M\,87 jet.

\vspace{-0.7cm}
\section{An Example:  Possible Jet and Cocoon Compositions}
\label{Scaling}
\vspace{-0.1cm}

In this section we explore an example of the possible internal state of the jet and the cocoon.  Recall that our models do not derive the density or pressure of either system, but only the specific internal energy of each.  Without definite information on the composition of the jet or cocoon plasmas, we cannot go any further.  However, we can gain insight by considering an example  in which both the jet and cocoon  contain a mix of relativistic and thermal plasmas, both scaled to the likely pressure in the M\,87 jet.

\vspace{-0.7cm}
\subsection{Composition:  Mixed Plasmas}
\label{composition}
\vspace{-0.1cm}

Let us consider a mixed plasma, which could describe the jet or the cocoon.  Let it contain ``thermal'' ions, with some temperature $T$ so that $P_{\rm th} = n_{\rm th} k T$. Let it also contain relativistic leptons, with $P_{\rm rel} = \mgamma n_{\rm rel} m_e c^2$, where $\mgamma \gg 1$ describes the effective temperature of the leptons, normalized to $m_e c^2$.  Note this differs from the bulk flow Lorentz factor of the plasma, which is much smaller.  The leptons inside the jet are observed by means of their synchrotron radiation. The ions could provide charge balance with the relativistic leptons, or could represent a separate, ion-electron population. For the cocoon, the thermal component could have its origins in the galactic ISM, and the relativistic component
could have its origins in the jet plasma, now mixed with the inner ISM and visible as the inner radio halo.

Our models show the plasma obeys $P = \Theta \rho c^2$, with $\Theta <
1$. From this,
\begin{equation}
P = P_{\rm rel} + P_{\rm th} = n_{\rm rel} \mgamma m_e c^2 + n_{\rm th} k T
= n_{\rm rel} \Theta  m_e c^2 + n_{\rm th} \Theta m_p c^2 ~.
\label{pressure_balance}
\end{equation}
We guess that the relativistic leptons are well separated in energy   from the thermal plasma.  For instance, we know the jet spectrum continues to rise down to $\sim 300$ MHz \citep{OEK2000}.  If this were a  low-frequency cutoff to the jet spectrum, and if the jet magnetic field  $\ltw 100 \mu$G (as in Section \ref{pressure_scaling}) the lepton  pressure would be dominated by particles at $\mgamma \gtw 300$, so
that $\mgamma \gg \Theta$. 

We normalize the ion temperature as
$\zeta = k T / m_pc^2$, with no constraints on $T$ except that $\zeta
< \Theta$ is required to solve Equation (\ref{pressure_balance}).
Solving that equation gives
\begin{equation}
{ P_{\rm rel} \over P_{\rm th} } \simeq { \Theta - \zeta \over \zeta} ~ ~; \qquad
{ n_{\rm rel} \over n_{\rm th} }\simeq { \Theta - \zeta \over  \mgamma} { m_p \over m_e}
\label{rel_th_ratios}
\end{equation}
At HST-1, where $\Theta_j \sim 0.04$, and $\Theta_c \sim 0.002$, the thermal plasma dominates the number density in both jet and cocoon (for any $\mgamma \gtw 100$).  The pressure ratio depends on the
unconstrained factor $\zeta$; but unless the thermal plasma is much hotter than the $T \sim 10^7$~K typical of the galactic ISM, both jet and cocoon pressures are dominated by the relativistic components.  At
knot A, where $\Theta_j \sim \Theta_c \sim 0.5$, the detailed numbers may change but the situation is probably the same:  the densities of the cocoon and jet plasmas are dominated by the thermal component, but their pressures are dominated by the relativistic component.

\vspace{-0.7cm}
\subsection{Pressure, Density and Magnetic Field}
\label{pressure_scaling}
\vspace{-0.1cm}

Up to this point our modeling has not depended on absolute  values for the density or the pressure but only on the ratio of   these two quantities.  We can estimate  the jet or cocoon  densities by choosing one more parameter, namely, the jet pressure. From Equation (\ref{pressure_balance}) we find that
\begin{equation}
n_{\rm rel}  = { \Theta - \zeta \over \Theta} { P \over (\mgamma - \zeta) m_e c^2} ~ ~; \qquad 
n_{\rm th} = { P \over \Theta m_p c^2} { \mgamma - \Theta \over \mgamma - \zeta}
\label{jet_densities}
\end{equation}
To estimate the pressure at our fiducial point, HST-1, we start with the minimum-pressure estimate from OHC, $P_{\rm min} \sim 2 \times 10^{-8}$~dyn~cm$^{-2}$.  This value will be reduced a bit by the $\delta^{-10/7}$ Doppler correction, but the true jet pressure must exceed $P_{\rm min} $.  We thus scale the pressure at HST-1 as $P_{j0} = 10^{-8}P_{0,8}$~dyn~cm$^{-2}$.  We guess that the relativistic leptons at HST-1 have $\gamma_{j, {\rm eff}} \sim 300$, as above. We simply scale the relativistic component of the cocoon to $\gamma_{{\rm
    eff},c} = 100 \gamma_{c,100}$, for lack of any better information about relativistic leptons in the cocoon.

Putting in numbers, we start at HST-1, with $\Theta_j \sim 0.04$ for the jet; this gives $n_{j,\rm rel} \simeq 4 \times 10^{-5}P_{0,8} $ cm$^{-3}$, and $n_{j,\rm th} \simeq 2 \times 10^{-4} P_{0,8}$ cm$^{-3}$.  For the cocoon at HST-1, with $\Theta_c \sim .002$, we get $n_{c,\rm th} \sim 3 \times 10^{-3}P_{0,8}$ cm$^{-3}$ and $n_{c,\rm
  rel} \simeq 1 \times 10^{-4}P_{0,8}/\gamma_{c,100} $ cm$^{-3}$. At knot A, our models say the pressure drops by a factor $3.5$ relative to that at HST-1. Both the jet and the cocoon have $\Theta_j \sim \Theta_c \sim 0.5$.  If $\gamma_{j, {\rm eff}} \propto \Theta_j$ holds for the jet, as above, we have $\gamma_{j, {\rm eff}} \sim 3800$ at knot A.  With these numbers, the jet densities are $n_{j,\rm rel} \sim 1 \times 10^{-6} P_{0,8}$ cm$^{-3}$, and $n_{j,\rm th} \sim 4 \times 10^{-6} P_{0,8}$ cm$^{-3}$.  For the cocoon, we have $n_{c,\rm th} \sim 4 \times
10^{-6}P_{0,8}$ cm$^{-3}$, and $n_{c,\rm rel} \simeq 4 \times 10^{-5}P_{0,8}/\gamma_{c,100} $ cm$^{-3}$ at knot A.  Thus, our assumed cocoon is not only hot, but also tenuous and overpressured, when compared to the
$n_{\rm th} \sim 0.1$ cm$^{-3}$ and $T \sim 10^7$~K typical of the X-ray loud ISM in the inner few kpc of the galaxy.  We discuss this cocoon further in Section \ref{Invisible_cocoon}.

Our models assume the plasma is only weakly magnetized, $P_B = B^2 / 8 \pi \ll P_j$.  Numerically, this gives $B \ll 500 P_8^{1/2}  {\mu G}$. Because $P_j \sim P_{\rm rel}$ in our example, our models must also have $P_B \ll P_{\rm rel}$. At first glance, this seems to contradict the results of Stawarz et al.\ (2005), who used
upper limits on inverse Compton emission, together with minimum-energy analysis of the radio power, to argue $P_B \gtw P_{rel}$ in knot A. We note, however, that different assumptions in such modeling (compare, e.g., OHC), as well as smaller assumed viewing angles (as in Section \ref{Jet_speed}) can accomodate a weaker field throughout the jet.  In addition, Stawarz et al.\ (2005) assumed the magnetic field and relativistic leptons coexist uniformly throughout the jet.  If, alternatively, the magnetic field  is enhanced at and beyond knot  A, or is strong only in localized structures (say a thin surface   layer around the jet, or the knots within the jet), the {\it
inner} jet may still have $P_B \ll P_{\rm rel}$ throughout most of its volume, in accord with our models.

\vspace{-0.7cm}
\subsection{Jet Power}
\vspace{-0.1cm}

We can also compute the total jet power (energy flux) in terms of the pressure at HST-1. In Section \ref{EC_jet} through Section \ref{EL_jet} we worked with the total energy flow down the jet (Equations \ref{EnergyCon} or \ref{two_phase_energy}).  However, when considering the power the jet can deposit in its surroundings, the
``usable'' energy (kinetic plus internal, omitting rest mass energy of the plasma particles) is more germane.  Thus, we write the jet power in the absence of significant magnetic energy as \citep{OEK2000}
\begin{equation}
\dot E_j = \pi r_j^2  \gamma_j^2 \beta_j c \left[ {( \gamma_j -1) \over \gamma_j}  \rho_j
  c^2 +{\Gamma_j \over \Gamma_j -1} P_j  \right]~.
\end{equation}
The first term in square brackets is the ``useful'' kinetic energy of the flow;  the second term is the internal energy.  This equation can also be written as 
\begin{equation}
\dot E_j = \pi r_j^2  \gamma_j^2 \beta_j c P_j { \Gamma_j \over \Gamma_j -1} 
\left[ {( \gamma_j -1) \over \gamma_j \chi_j} + 1 \right]~.
\label{jet_power}
\end{equation}
Because our jet models are mildly relativistic ($\gamma_j \sim$ a few), and have $\chi_j \sim O(1)$, both kinetic and internal energies contribute significantly to the total jet power. 

Let us evaluate Equation (\ref{jet_power}) at HST-1, where $r_j \simeq 1.3 \times 10^{19}$~cm and  $P_{j0} = 10^{-8}P_{0,8}~$dyn cm$^{-2}$. For our fast jet, with  $\gamma_{j0} = 7.5, \chi_{j0} = 0.11$ and
$\Gamma_{j0} = 1.63$, we get $\dot E_j \simeq 2.0~P_{0,8} \times 10^{44}$ erg s$^{-1}$. For our slow jet, with $\gamma_{j0} = 4.4, \chi_{j0} = 0.08$ and $\Gamma_{j0} = 1.64$, we get  $\dot E_j \simeq 8.2~P_{0,8} \times 10^{43}$ erg s$^{-1}$. These values are more or less consistent with estimates for the M\,87 jet power in the literature, which range from $\dot E_j \sim 2 \times 10^{43}$ erg s$^{-1}$, required to drive
the weak shocks associated with the present jet activity \citep{Form05}, to $\dot E_j \gtrsim 10^{44}$ erg s$^{-1}$, based either on the energy required to expand the inner lobes into the surrounding ISM over a $\sim$~1~Myr lifetime \citep{BB96}, or on the minimum jet power required at knot A as derived from synchrotron
theory (as in Section \ref{pressure}; \citet{OEK2000}).

\vspace{-0.7cm}
\section{A Pseudo-Synchrotron Image of Our Model Jet}
\label{Pseudo_Synch}
\vspace{-0.1cm}

In this section we illustrate the effects of the KH modes on the appearance of a jet by producing a model image which we compare to the real M\,87 jet.  We choose the fast-intermediate-case jet model as being the most likely scenario for the real jet. We compute the intrinsic pressure and
velocity structure along a jet with a mix of helical, elliptical, triangular and rectangular surface modes, chosen so that the elliptical surface mode dominates the structure.  We use these results to generate a pseudo-synchrotron intensity image that includes all differential Doppler boosting and light travel time effects associated with the relativistically moving pressure and velocity structures.  We view this model jet as being ``representative'' of a jet like that in M\,87, but have not attempted to create an exact match to the real jet. 

\vspace{-0.7cm}
\subsection{Pressure and Velocity Structure of the KH Modes}
\vspace{-0.1cm}

We compute the pressure and velocity flow fields associated with relativistically moving normal mode structures in the limit of weakly non-linear, adiabatic compressions. To do this, we need expressions for the spatial structures of the perturbations associated with each KH mode.  In the linear regime the displacement amplitude of the jet surface associated with a normal mode obeys
\begin{equation}
A(z)  = A_{j0}~\exp \left[ \int_{z_{j0}}^{z} |k_i(z)|  dz \right]~,
\end{equation}
where $A_{j0}$ is the amplitude at $z_{j0}$ of a sinusoidal displacement to the jet's surface, $k_i(z)$ is the spatial growth rate for that mode (Figure \ref{Spatial_growth}), and $\int |k_i(z)|dz$ is ``the number of growth lengths along the jet.'' We use the expressions found in Section 3 of \citet{H00} to calculate the
pressure and velocity fluctuations. These grow more slowly than the surface displacement, according to
$A(z)/r_j(z)$ rather than $A(z)$, because the displacement amplitude as a fraction of the jet's
radius governs the magnitude of velocity and pressure fluctuations.  Thus, for constant jet expansion,
\begin{equation}
{A(z) \over r_j(z)} = {A_{j0} \over r_{j0}}~{\exp \left[ \int_{z_{j0}}^{z} k_i(z) dz \right] \over
\left[ 1 + (z - z_{j0})\psi/r_{j0} \right]}~,
\label{spatial_growth}
\end{equation}
where  $r_j = r_{j0} + (z - z_{j0})\psi$ and $\psi$ is the half opening angle of the jet.  The small opening angles in our models, $\psi \sim 0.0523 \sin\theta$ (where $\theta = 15\arcdeg$ or $25\arcdeg$ is the viewing angle), are significantly less than the relativistic Mach angle, $ \sim (\gamma_j M_j)^{-1}$ at HST-1, as is required for self-consistency. The spatial growth length, $\ell = |k_i|^{-1}$, for the elliptical surface mode between HST-1 and knot A
when rescaled to the jet's radius at HST-1 increases from $\ell \sim 10 r_{j0}$ to $\ell \sim 20 r_{j0}$.  For a typical value of $\ell \sim 15 r_{j0} \sim 60$~pc, pressure and velocity fluctuations  grow by an order of magnitude in $\sim 150$~pc.

We have explored a number of different mixes  for four lower-order KH modes (helical, elliptical, triangular, rectangular) using spatial growth rates for the fast-intermediate-case jet model to evaluate Equation (\ref{spatial_growth}).  Our goal is not to reproduce the M\,87 jet exactly but to generate the most educational example.  To do this we have arbitrarily capped helical and elliptical modes at an identical amplitude so that the pressure fluctuation induced by the helical mode is about 50\% of that induced by the elliptical mode. We also cap the triangular and rectangular modes at a smaller amplitude, to create a pressure fluctuation also about 50\% of that induced by the elliptical mode. With these amplitudes the combined modes produce a maximum pressure fluctuation comparable to the pressure field in the unperturbed jet, i.e., $\Delta P \lesssim P_j$.  This is consistent with simulations \citep{H98,A01,HH03} which show that the linearized equations satisfactorily model the pressure and velocity structure in the simulations at this fluctuation level.

\begin{figure}[h!]
\vspace{10.7cm}
\includegraphics{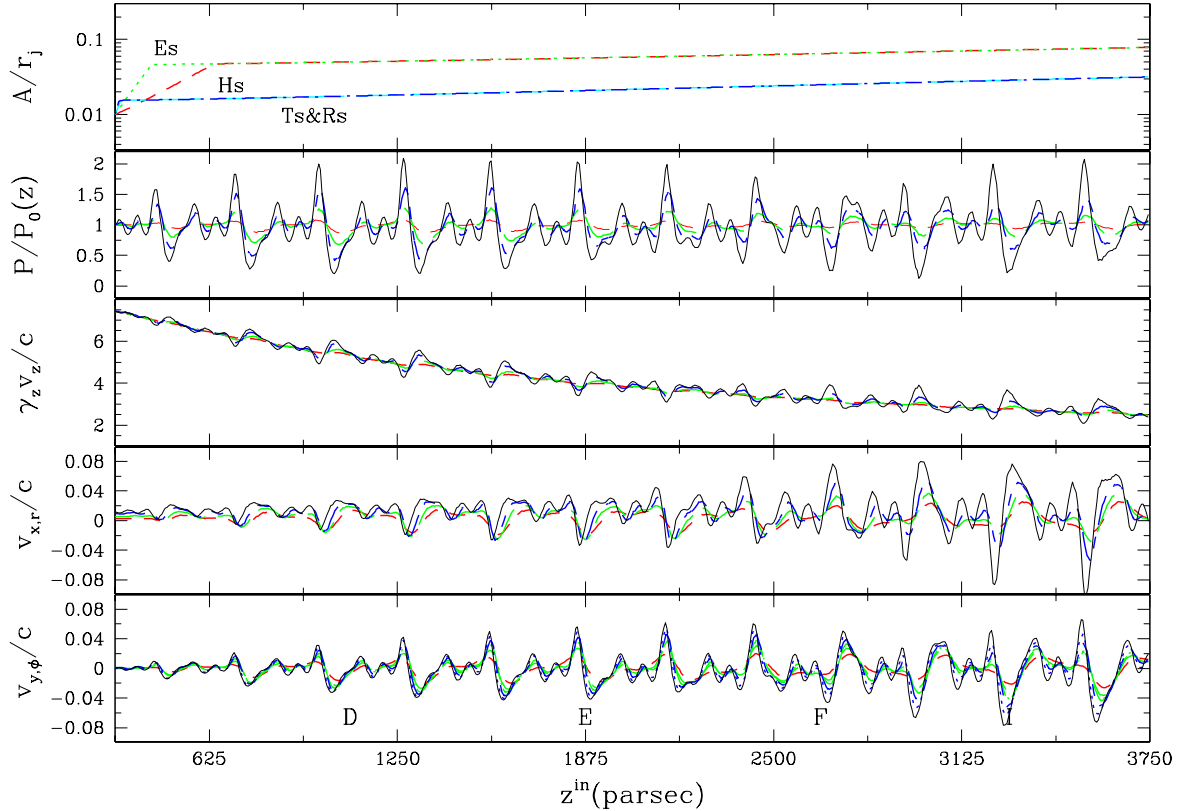}
\caption{\footnotesize \baselineskip 11pt Top panel shows the  amplitudes, $A/r_j$, for (Hs) helical, (Es)
elliptical, (Ts) triangular \& (Rs) rectangular surface modes used in the  lower panels to compute intrinsic pressure and velocity fluctuations. Quantities are plotted against  $z^{\rm in}$, the intrinsic distance down the jet.  In pressure and velocity panels 1D cuts are shown at: $r/r_j = x / r_j = 0.22$ (red dashed
lines), $0.44$ (green dashed lines), $0.66$ (blue dashed lines), $0.88$ (black solid lines).}
\label{Fluc_amplitudes}
\vspace{-0.2cm}
\end{figure}

Our results are shown in Figure \ref{Fluc_amplitudes}. The top panel indicates how rapidly a small initial perturbation with amplitude $A(z)/r_j(z) = 0.01$ at HST-1 would grow to our amplitude caps on the M\,87 jet for frequency-resonant helical, elliptical, triangular and rectangular surface modes. The lower panels of Figure
\ref{Fluc_amplitudes} show one-dimensional pressure and velocity cuts, made in a Cartesian coordinate system at $y = 0$ and $x/r_j =$ 0.22, 0.44, 0.66, 0.88. Thus, $v_x = v_r$ is a ``radial'' velocity component in cylindrical coordinates and $v_y = v_\phi$ is a toroidal velocity component.  

The results illustrate the complex structure that can result from multiple mode superpositions. The dominant wavelength of the pressure fluctuation, increasing along the jet but typically $\sim 300$ pc, is the wavelength of the elliptical surface mode corresponding to a $180\arcdeg$ rotation of the elliptical distortion to the jet's cross section. Variation in the fluctuations comes from beating between the different wavelengths and wave speeds associated with the other KH modes.  The strongest pressure fluctuations occur near the jet's surface.  The velocity fluctuation in the axial direction, $\gamma_z v_z$, is small. The modest increase in ``radial'' velocity, $v_r$, towards the jet surface indicates jet expansion. There is more fluctuation in radial $v_r$ and toroidal $v_{\phi}$ motion near the jet's surface.

\vspace{-0.7cm}
\subsection{The Pseudo-Synchrotron Model Image}
\vspace{-0.1cm}

In order to image our model jet, we  create a three-dimensional, Cartesian, data cube which  we use in a modified version of RADIO \citep{CNB89}.  Each grid element in the data cube is derived from the density, pressure and velocity values we calculate throughout the jet, as in  Figure \ref{Fluc_amplitudes}.  These values are suitably corrected for all light-travel-time effects associated with relativistically moving structures seen at a viewing angle of 15\arcdeg. We use a pseudo-synchrotron emissivity, based on standard synchrotron theory for a power-law distribution of relativistic lepton energies, with Doppler boosting included:
\begin{equation}
\epsilon_{\nu}  \propto \rho_j^{1 - 2 \alpha}P_j^{2\alpha}B^{1 + \alpha}\delta^{2 + \alpha} \nu^{- \alpha}~.
\label{pseudo_synch}
\end{equation}
Here, $\epsilon_{\nu}$ is the volume emissivity per Hz; the spectral index $\alpha = 0.55$  (appropriate for the M\,87 jet in the radio band; Frazer Owen, private communication 2009) is taken as constant.  This expression assumes the leptons are coupled to the total mass density in the jet, are subject only to adiabatic compression, and ignores radiative losses and {\it in situ} energization.  \citet{JRE1999} have shown that  Equation (\ref{pseudo_synch}) provides an acceptable image of jet structure in many cases when the
relativistic particles are not tracked explicitly.  We assume the field strength obeys $B \propto \rho_j^{2/3}$, for a passive, disordered magnetic field.  We take $P_{j} (z) \propto \rho_{j}(z)^{0.325}$, to describe the pressure decline in our intermediate-case model jet, and on average the emissivity drops according to  $\epsilon_{\nu} (z) \propto \rho_{j}(z)^{1.3} \propto P_j(z)^{4.0}$. Adiabatic pressure fluctuations obey $P \propto \rho^{\Gamma}$ around the local values, where $4/3 \le \Gamma \le 5/3$ (equation \ref{G_approx}).  Thus, a factor $\sim 1.5\!-\!2$ pressure increase towards the outer edge of the jet in a filament (Figure \ref{Fluc_amplitudes}) increases the local emissivity, relative to the mean value,  by $\epsilon_{\nu} \propto P_j^{(1.1\Gamma +0.93)/\Gamma}
\sim P_j^{1.7} - P_j^{1.8}$, thus by a  factor $\sim 2\!-\!3$.

\begin{figure}[h!]
\vspace{-6.50cm}
{\center 
\includegraphics[width=1.0\textwidth]{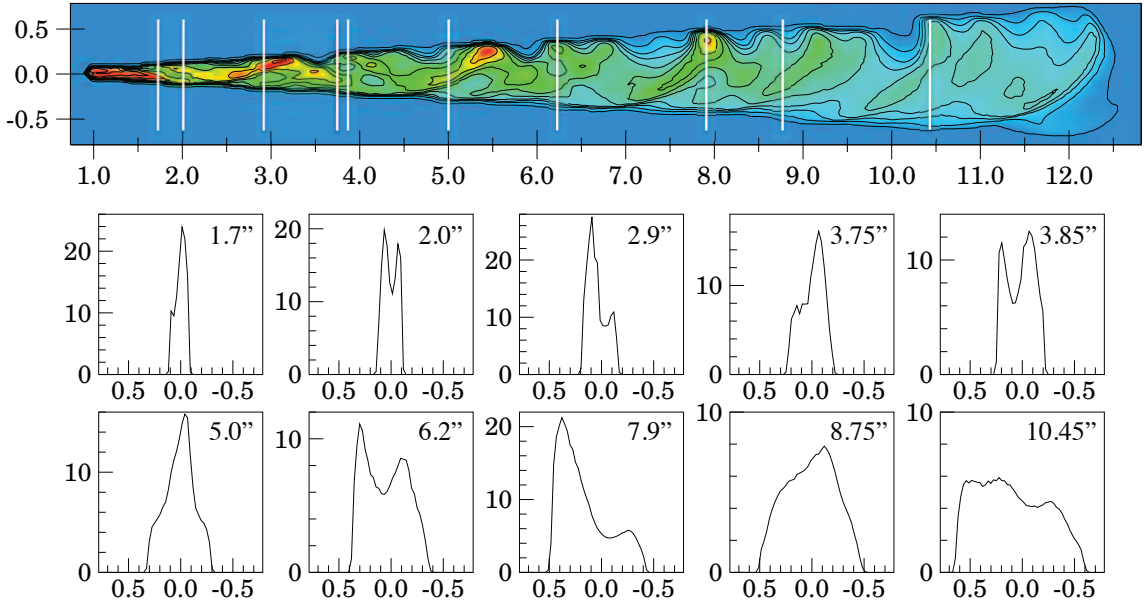}
\vspace{-7.25cm}
\caption{\footnotesize \baselineskip 11pt Pseudo-synchrotron intensity image (upper).  The axes are in units of arcsec and contours are at the levels 0.1, 1.25, 2.5, 5, 7, 10,  14, 20, 28, and  40 intensity units per pixel. 
The resolution is $\Delta = r_{j0}/2 = 0.0262\arcsec$, a pixel is a $\Delta \times \Delta$ square, and four pixels span the jet diameter at 1\arcsec.  Transverse intensity profiles (lower), taken across the jet; axes for these profiles are (vertical), intensity units per pixel and (horizontal) arcsec across the jet.  Each profile is labelled with its position along the jet in arcsec. }
\label{Pseudo_image}}
\end{figure}

We present our model image in Figure \ref{Pseudo_image}.  
The upper panel shows the basic image, with intensity levels indicated by color scale and contour levels.  The lower panels show slices across the model jet at ten locations along the jet.  At a fixed spatial location, the filaments would rotate temporally in the clockwise direction when viewed towards the central engine by an observer, corresponding to a clockwise spatial twist of the filaments when viewed from the central engine outwards. If the opposite twist were used the transverse asymmetry seen in Figure \ref{Pseudo_image} would be reversed. The importance of relativistic aberration and jet orientation is illustrated by comparing Figure \ref{illustrate_KH} to  Figure \ref{Pseudo_image}.  The jet in Figure \ref{illustrate_KH}, chosen to illustrate the intrinsic nature of the KH elliptial mode,  is at rest in the plane of the sky.  The jet in Figure \ref{Pseudo_image} has a similar intrinsic pressure structure, dominated by the elliptical mode, but it is moving relativistically and sits at $15\arcdeg$ to the line of sight. The differences in its appearance, compared to the jet  in Figure \ref{illustrate_KH}, are due to its speed, orientation and the presence of additional dynamically significant KH modes.

\vspace{-0.7cm}
\subsection{Comparing the Model Image to the Real Image}
\vspace{-0.1cm}

While our model jet image is not tuned to reproduce the M\,87 jet, we gain considerable insight by comparing our model jet to the 15-GHz image of the M\,87 jet in Figure \ref{Jet_inner} of this paper
(also Figures 5, 6 and 7 of OHC), and the NUV image of \citet{Madrid2007}.  Our model image shows some of the features seen in the real image, but differs revealingly in others. 

Along the model jet the average intensity drops by a factor $\sim 10$ between $1\arcsec$ and $12\arcsec$.  This drop is much less than might be expected for an expanding jet which does not decelerate.  In a constant-speed jet, expansion by a factor $\sim 12$  between $1\arcsec$ and $12\arcsec$ would cause the density to drop by a factor $\sim 140$, and the pressure to drop adiabatically by $\sim 4000$, leading to a drastic decline in the synchrotron emissivity.  Our intensity decline is still somewhat more than in the radio image of the real M\,87 jet;  in Figure \ref{Jet_inner} the jet intensity measured between the bright knots drops by a factor $\sim 2$ between HST-1 and knot A. Possible reasons for this difference not accounted for in Equation (26) include a slower decline with expansion in an ordered magnetic field and continuous {\it in situ} energization near to the jet's surface decoupling the radiating leptons from the dynamical density and pressure field.  These combined effects could lead to the observed slower overall intensity decline and to the observed limb brightening which requires an emissivity from the outer $20\%$ of the jet $\sim 5$ times higher than that of
the inner part of the jet (see OHC).

Our model image and the accompanying slices convey an overall impression of filaments wrapping around the jet.  
The twisted filaments can also be traced in the slices in Figure \ref{Pseudo_image}. The four slices from $1.7\arcsec$ to $3.75\arcsec$ show winding filaments which brighten alternately at the top and bottom edge of
the jet, but followed by rapid spatial change from $3.75\arcsec$ to $3.85\arcsec$. Various different transverse structures are illustrated by the slices from $5.0\arcsec$ to $10.45\arcsec$ including an overall decline in intensity and lack of sharp filament structure beyond $8.0\arcsec$. For comparison, filaments in the real jet are narrower than those in our image and cross the jet at a more oblique angle, e.g., profiles across the real jet (in OHC), from $\sim 6\arcsec$ to $ \sim 10.5\arcsec$. 
 
The model image  is much more asymmetric past $\sim 4\arcsec$ than the real jet, with the upper half of the model image being $\sim 2-3$  times brighter than the lower half. In large part this is a result of one elliptical filament being much brighter than the other. The brighter elliptical filament can easily be traced in the intensity image in Figure \ref{Pseudo_image} as it has produced bright spots along the jet top.  This 
filament completes a $ 360\arcdeg$ wrap between $1.25-3.25\arcsec$, again between $3.25\!-\!5.5\arcsec$, yet again between $5.5\!-\!8\arcsec$, and finally again between $8\!-\!11\arcsec$.  The filament twists around the jet with wavelength varying from $2\arcsec \!-\! 3\arcsec$ along the jet in agreement with the results from LHE. The second elliptical filament comes over the jet top about midway between the bright spots associated with the first filament and no bright spots are apparent. The difference between the two filaments occurs because the pressure structure accompanying the included helical mode adds pressure to one elliptical filament and subtracts from the other.  In the real jet the two filaments are comparable and no bright spots are seen along the top edge.

The model image leads us to conclude that the helical mode on the real jet must be at very low amplitude and dynamically insignificant inside knot A in order for both filaments to be comparable. The model image requires a somewhat lower elliptical amplitude in order to reduce the bright spots along the jet top that are not seen on the real jet.  In our model image low-level triangular and rectangular modes add some complexity, as expected from Figure \ref{Fluc_amplitudes}, and make it more similar to the real jet. Note that our model emissivity is assumed to have no radial dependence other than that caused by the KH instability. The diminished emissivity in the inner 80\% of the real jet would also lead to differences between the model and real jet images. 

Even with elimination of the helical mode and reduced amplitude for the elliptical mode, our model image would still show an asymmetry.  The asymmetry is not due to  flow around the helix; Figure \ref{Fluc_amplitudes} shows that   transverse velocity fluctuations, $v_{r,\phi} \ltw 0.1c$, are too  small to produce significant differential Doppler boosting. Rather,   the asymmetry accompanies any helical pattern projected on the sky  plane. The amount of asymmetry depends on the speed of the helix and  its radius relative to its wavelength. The helix  appears symmetric if viewed close to the critical angle, $\theta_c \equiv  \cos^{-1} \beta_w$, but becomes progressively more asymmetric as the viewing angle moves away from $\theta_c$ (Perucho et al., in preparation). In the  model image the asymmetry is minimal in the inner $4\arcsec$ where  the filament winds around the jet surface and $r_j \ll \lambda^{ob}/2\pi$. However, with our  viewing angle $\theta \ll \theta_c$, the asymmetry becomes evident at larger distance when, $r_j \lesssim \lambda^{ob}/2\pi$ and  cusps appear along the jet top beginning at about $8\arcsec$.  There  is some indication of this behavior on the real jet beyond knot  F. At about the position of knot A our helical filaments would loop  back on themselves at the top of the jet and this behavior might be indicated by the complicated structure of knot A in the real jet (see Figure 2).

The real jet contains bright knots inside the jet edges, which are not associated with filament crossings, and appear to be features internal to the jet rather than on its surface. The knots in the real jet are 
enhanced in the radio image by a similar factor of $\sim 3-5$ compared to the fainter interknot background, but are relatively brighter and more compact in the optical image.  Our model image shows that these bright knots in the real jet cannot be the result of twisting KH modes. The apparent knot location inside the real jet, and the fact that details of knot structure change with the observing band, suggest that an additional mechanism is operating in the jet interior.
 
\vspace{-0.7cm}
\section{Summary}
\label{Summary}
\vspace{-0.1cm}

We have presented a model in which the dual, helically twisted filaments in the inner M\,87 jet are caused by the Kelvin-Helmholtz (KH) instability.  We based our model on the qualitative similarity between the observed
filaments and the likely development -- linear growth and nonlinear saturation -- of the KH elliptical surface mode. The model was developed in three stages.  We first constructed baseline, steady-state jet models.  We then carried out a KH analysis of the baseline models.  Finally, we  created a pseudo-synchrotron image of KH modes in a representative model jet, and compared it to the real jet in M\,87.  In this section we summarize the key results of each step of our analysis.

\vspace{-0.7cm}
\subsection{A Steady-State Jet Model}
\label{SSJM}
\vspace{-0.1cm}

Proper motions of bright features in the jet are consistent with a faster jet flow moving through a slower wave pattern.  Based on these data, we  modelled the (unperturbed) inner jet as a weakly magnetized, relativistic,  steady-state flow, oriented at $15^{\circ} - 25^{\circ}$ to the line of sight, and found the following.

{\it The jet flow decelerates  between HST-1 and knot A}. It begins with $\gamma_{j0} \sim 4 - 8$ at HST-1, and slows to $\gamma_j \sim 2 - 3$ by knot A.  This deceleration is (almost) directly observed in the superluminal
proper motions of the faster bits of the bright knots.  The range of $\gamma_j$ values reflects variation in reported proper motions and uncertainty in the viewing angle.  

{\it The wave (pattern) accelerates along the same part of the jet}. This is required by the observed increase in filament wavelength along the jet (from LHE), and is consistent with subluminal proper motions detected in slower bits of the bright knots.

{\it The jet plasma is internally hot but subrelativistic}.  If the jet conserves mass and energy as it
decelerates, its internal energy must be subrelativistic with $P_j \sim .04 \rho_j c^2$ at HST-1 and heating to $P_j \sim 0.5 \rho_j c^2$ by knot A.  A modest ($\ltw 20\%$) loss of energy from the jet flow to a surrounding cocoon does not significantly change this result.  Thus, although the jet is detected by means of the highly relativistic leptons ($P_{\rm rel} \gg \rho_{\rm rel} c^2$) responsible for its synchrotron radiation, its internal energy is dominated by a substantially ($P_{\rm th} \lesssim \rho_{\rm th} c^2$) cooler component.  

\vspace{-0.7cm}
\subsection{KH Stability Analysis}
\label{KHSA}   
\vspace{-0.1cm}

We analyzed the KH instability of our steady-state jet models. Our analysis requires that the jet be in pressure balance with a surrounding ``cocoon''.  We determined what conditions must exist in this cocoon in order to account for the observed filaments in terms of the KH elliptical surface mode without any assumptions about how close to resonance the KH elliptical mode might be. Our results from this analysis are as follows.

{\it The jet is surrounded by a hot cocoon.}  The cocoon must be denser and cooler than the jet, but still quite hot compared to the galactic ISM. At HST-1 the cocoon must have $P_c \sim 0.002 - 0.003 \rho_c c^2$; its mass density must be $\sim 10 - 20$ times that of the jet.  By knot A it must have $P_c \sim 0.4 - 0.7 \rho_c c^2$, and its mass density must be comparable to that of the jet. These results are not derived from thermal modeling of the cocoon, but rather are required (through the solution of the KH dispersion relation) in order to match the increase in filament wavelength observed along the jet.

{\it The filaments are KH instabilities close to resonance.}  The wavelength of the filaments and the estimated wave speed  are found to be close to the resonant (fastest growing) wavelength expected for the KH instability under the conditions we derived for the jet and the cocoon. This was not assumed a  priori. Because we expect random perturbations to couple to the system's resonant frequency, we take this as circumstantial evidence supporting our hypothesis that the filaments in the M\,87 jet are produced by KH instability.

{\it The filaments are saturated KH elliptical modes}.  We determined growth rates for the KH modes, and verified that they are rapid enough for the observed filaments  to have developed by the position of HST-1.  Because
our model is based on a linear analysis, we cannot analytically predict the level at which the KH modes will saturate.  However, we suggest the elliptical surface mode is responsible for the filaments, because that mode is the most likely to dominate in the nonlinear stage without disrupting the jet.  The lack of any detectable helical twisting between the core and knot A requires that any helical perturbation to the jet be very small.

\vspace{-0.7cm}
\subsection{Synchrotron Image of the Model Jet}
\label{SIMJ}
\vspace{-0.1cm}

We created a pseudo-synchrotron image of one of our fast-jet models, which was modified to include helical through rectangular KH surface wave modes. The image took light-travel-time and relativistic wave motion into account for our assumed viewing angle.   We  did not attempt to match specific features in the real jet, but comparison between our model image and the real image revealed the following.

{\it The surface brightness along the model jet  declines more rapidly than in the real jet but much less so than in adiabatic expansion.}  This is directly due to our assumption that the jet plasma decelerates between HST-1 and knot A and energy flux is conserved. The deceleration and conversion of flow to internal energy offsets the radial expansion of the jet, and keeps the psuedo-synchrotron emissivity high.  The less rapid decline in the real jet indicates that additional in situ processes must be operating.

{\it The twisted filament structure in the model jet differs from that in the real jet}.  The filaments in the real jet appear more symmetric and are narrower than those in the model jet. On average the model jet filaments are brighter along one edge due to the angle at which the helically twisted filaments are viewed. One of the two filaments in the model jet has been overly enhanced by the inclusion of the helical surface mode. This indicates that any helical perturbation to the jet inside knot A must be very small or nonexistent.   The real jet is more limb brightened, especially in the radio, than is the model jet. This suggests that additional processes are involved in the appearance of the real jet on the edge of the flow.

{\it The bright knots in the real jet are not produced in the model jet.}  The bright knots in the model jet that lie along  one edge are due to the angle at which the helically twisted filaments are viewed and to the inclusion of the helical surface mode.   The bright knots in the real jet tend to have a more complex structure than those in the model jet, and also tend to lie inside the jet.  This indicates that additional processes not associated with KH instability are involved in the real jet inside the flow. 

\vspace{-0.7cm}
\section{Discussion}
\label{Discussion}
\vspace{-0.1cm}

How might our model fit within the larger picture of what is known about the inner M\,87 jet and its environment?  Specifically, let us consider the presence of a hot cocoon, the origin of the knots, and the disruption of the jet past knot A.

\vspace{-0.7cm}
\subsection{Is There an Invisible Cocoon?}
\label{Invisible_cocoon}
\vspace{-0.1cm}

A key part of our analysis is the existence of a hot, tenuous, high-pressure cocoon surrounding the inner jet in M\,87.  Such a cocoon must exist if the filaments in the jet are caused by the KH instability. Specifically, our models require the cocoon to satisfy $P_c \sim 0.002 \!-\! 0.5 \rho_c c^2$, which is substantially hotter than the $T \sim 10^7$K typical of the inner ISM.  Our specific example, in Section \ref{Scaling}, suggests a cocoon density $\sim 10 $ times lower than the inner ISM at HST-1, and $\sim 10^3$ times lower than the inner ISM at knot A.

At first glance this requirement seems to contradict the data, because no such cocoon has been detected; but it may be hard to see.  Any X-ray detection of the cocoon would be indirect; it is too tenuous, and possibly too hot, to be X-ray bright.  It could appear as a ``hole'' in an X-ray image, but the wealth of loops, filaments and
cavities in the inner X-ray region \citep{Form05} could easily hide such a structure. Direct detection of the cocoon in the radio is equally difficult, because uncertainties in relativistic particle energy and magnetic field in the cocoon make predictions of its radio brightness elusive. Figure \ref{Jet_large} shows that the observed radio halo is relatively faint around the inner jet. This could mean that the region immediately around the jet is more tenuous than elsewhere in the radio halo, or has a weaker magnetic field, or that the radiating leptons have suffered significant losses.  These possibilities are consistent with our hypothetical cocoon, but by no means confirm its existence. A hint of a low-density cocoon may come from the Faraday rotation, which is substantially lower along the jet than elsewhere in the inner radio lobes \citet{OEK1990}.  While this could be simply due to projection if the jet lies in front of most of the inner ISM, it could also indicate a low-density region local to the jet.

Another possible concern comes from the requirement that the cocoon be at a higher pressure than the local ISM.  Interpreting the filaments as being KH generated requires the cocoon to be in pressure balance with the jet; but the minimum pressure of the jet, $P_{\rm min}$, from synchrotron analysis, is higher than the typical pressure of the inner ISM.  It follows that the jet-cocoon system cannot be in a steady state; it must be dynamic, and expanding outwards into the local ISM.  This seems consistent with the evidence that the ISM immediately surrounding the jet is disordered and turbulent (as in Section \ref{flow_obs}).  The jet itself is clearly the driver for the inner ISM.  It is depositing mass and energy directly into the inner radio halo (as in Figure \ref{Jet_large}) and our models suggest it can be shedding significant energy into its immediate cocoon. 

Yet another concern might be the origin of such a cocoon and why it is so much hotter around knot A than at HST-1.  While our model does not address this issue, we note that the relativistic plasma from the jet is likely to have mixed with the ambient ISM.  Such mixing is directly suggested by the existence of the radio halo  (as in Figure \ref{Jet_large}), and is consistent with our example of jet and cocoon composition in Section \ref{Scaling}.  We speculate that relativistic plasma  from the jet is mixed with the ambient plasma as the jet propagates outward through the galactic ISM.  Over time, the changing spatial location of the primary mixing and heating region might lead to the required temperature and sound speed increase in the cocoon from HST-1 to knot A.

\vspace{-0.7cm}
\subsection{What Causes Knots and Limb Brightening?}
\label{knotsand limbs}
\vspace{-0.1cm}

There are two important differences between the real jet and our model image.  The real jet displays the well-known bright knots, which appear to be internal to the jet (as seen in projection) and not coincident with filament crossings.  Our model jet does not have such complex bright knots. In addition, the real jet is limb brightened when seen in the radio, but  our model jet is not. 

These differences between the real jet and the model suggest that the jet plasma is being energized {\it in situ}, in the bright knots as well as at the edge of the jet.  This is not a new argument.  The radiative lifetimes of relativistic leptons emitting X-ray synchrotron emission are too short for the particles to survive the transit from the core out to knot A.  This result comes directly from the strong galactic starlight \citep{SSO05}, and holds for any plausible magnetic field in the jet.  It follows that the X-ray-loud leptons must undergo local reacceleration throughout the jet.  We think it likely that lower-energy particles radiating in optical and radio, are also locally energized by the same processes.  Spectral behavior  along the jet (e.g. Perlman \& Wilson 2005, also F. Owen, private
  communication 2009) requires that the acceleration mechanism, whatever it  is, maintain an approximately constant lepton spectrum throughout   the jet.

One obvious possibility for {\it in situ} energization is the KH instability itself.  We showed in Section \ref{intermediate_jet} that the wave speed is supersonic throughout the jet.  Thus, the twisted filaments  provide a
continuous pair of twisted shocks, and possible particle acceleration near the jet surface,  from HST-1 to knot I. In addition, velocity shear at the edge of the jet may amplify the magnetic fields there, locally enhancing the synchrotron emissivity close to the edge.  If this is the case, the weaker limb brightening in the optical, as compared to the radio, must be due to a relatively low, high-energy cutoff of the particle distribution near the jet surface.

However, edge effects are not likely to be the full answer.  In general, the knots appear located within the jet, not close to its edge (although the HST-1 flaring region was at the jet edge), and contain compact, bright emission regions when seen in the optical or NUV.  Thus, the data strongly suggest that energization internal to the jet is also an important part of the picture.  This is unsurprising, because the jet is supersonic from HST-1 to knot I, and very likely to contain internal shocks.  Shocks could arise because the jet flow is unsteady; this is evident from the variability of the knots, e.g., \citet{BSM99}.  Shocks could also arise as the jet propagates through the turbulent ISM in the galactic core.  Particle acceleration at internal shocks could very plausibly create the bright knots seen in the radio, optical and X-ray. 

\vspace{-0.7cm}
\subsection{What Happens Past Knot A?}
\label{knotA}
\vspace{-0.1cm}

We can use our model to speculate on the behavior of the jet downstream of knot A. Radio images (see Figure \ref{Jet_large}) show the jet narrows, and its centerline oscillates, between knots A and C. Our models find the 
jet is hot and relatively slow at knot A, with sound speed $a_j \sim 0.5 c$ and Lorentz factor  $\gamma_j \gtrsim 2.5$. If knot A is a shock, the sound speed would be higher and the Lorentz factor would be lower immediately post-shock. In a linear analysis, at a fixed pressure fluctuation,  the transverse fluid displacement associated with a KH mode is larger on a weakly supersonic, weakly relativistic jet than it is on a highly supersonic and highly relativistic jet \citep{H00}. Thus, the narrowing of the jet past knot A where the flow could be weakly supersonic and weakly relativistic might be associated with significant elliptical distortion of the jet's cross section associated with the elliptical surface mode of the KH instability.

In addition, the highly collimated jet flow disrupts significantly downstream of knot C.  As we noted in Section \ref{KH_intro}, the KH helical mode does not saturate but grows in amplitude until the jet decollimates.  We speculate that a slowly growing, long-wavelength helical mode is what disrupts the jet downstream of knot A.  To test this idea, even with a discontinuity at knot A, we can estimate the downstream development of the KH instability in terms of flow conditions at knot A.  On a weakly supersonic jet,  KH stability analysis shows that the supersonic resonance moves to very high frequencies and very short wavelengths.  In this situation the development of KH surface modes can be estimated from the solution of Equation (\ref{low_freq_DR}), using $ \omega_H  =  \omega_E/2 $. Our models at knot A (e.g., Table 2) have $\omega r_j/u_j \sim 1$, also $\rho_j / \rho_c \sim \chi_j / \chi_c \sim 1$, $\gamma_c \sim 1$ and $\gamma_j \ltw 2$.  With these numbers, the growth length for the helical surface mode, $\ell(\omega_H) =  | k_i(\omega_H)|^{-1}$, is no more than a few jet radii.  Noting that $r_j \sim 50$ pc at knot A,  a helical perturbation there will grow by an order of magnitude in $\sim 200\!-\!300$ pc.  For comparison, the distance from knot A to knot C is $\sim 1.2$~kpc  with the jet at a $15\arcdeg$ viewing angle. Thus,  our results suggest that destabilization and decollimation of the M\,87 jet beyond knot A is determined by conditions in the jet before it reaches that point.  With minimal energy flux lost from the jet, the decelerating jet flow must heat up;  the cocoon must also heat in order to account for the observed helical filaments.  These conditions lead to rapid growth of the KH helical mode, and consequent jet disruption, past knot A. 

Our goal in this paper has been to uncover properties of the M\,87 jet and its surroundings by assuming the KH instability is responsible for the twisted filaments seen in the jet.  It is illuminating to compare our conclusions here to those of a  similar analysis of the 3C\,120 jet \citep{HWG2005}. Bright knots and  inter-knot emission in the 3C\,120 jet on scales of hundreds of pc, suggest conical jet flow and a more slowly moving short wavelength helically twisted filament which is consistent with a short wavelength helical surface mode.  The observed behavior of this mode in 3C\,120 requires a cooling, modestly accelerating flow along the expanding jet.  Under these conditions, \citet{HWG2005} argued that the 3C\,120  jet is stabilized to KH disruption by saturation and the declining growth rate of the short-wavelength mode along the  jet. Their result explains how the  3C\,120 jet can  remain highly collimated to hundreds of kpc from the central engine.  By contrast, in this paper we find a different  result for  the M\,87 jet.  We suggest that heating of the M\,87 jet  as it decelerates allows the KH helical mode to grow significantly downstream of knot A, leading  to jet disruption only a few kpc from the central engine.

\begin{acknowledgements}
 We thank Andrei Lobanov for his analysis of the M\,87 jet filaments which provided the initial motivation for this study, and Frazer Owen for providing the 15 GHz VLA images and unpublished results on the radio-band spectral index of the jet.  P. Hardee acknowledges support from NSF award AST-0908010 and NASA award NNX08AG83G to the University of Alabama.
\end{acknowledgements}

\vspace{-0.7cm}
\begin{appendix}
\section{Propagation and Growth of Unstable KH Modes}
\vspace{-0.1cm}

The propagation and growth of the normal modes of the KH instability depend on the jet and cocoon speeds,  the thermal state of the jet and the cocoon,  the rate of jet expansion, and  the frequency of the wave relative to a resonant frequency $\omega^*$ and wavelength $\lambda^* = 2 \pi v^*_w/\omega^*$ at which the growth rate is a maximum.  In this Appendix we review the basic structure of our KH analysis and analytic approximations for mode growth below, at, and above the resonant frequency.  We follow  \citet{H07} and refer the reader to that reference for more detail.

In cylindrical geometry (with jet flow along the $z$-axis) a random perturbation to the
velocity, pressure or density can be described in terms of Fourier components  of the form
\begin{equation}
f (r, \phi , z , t ) =  f (r) e^{ [i(k z \pm n \phi -\omega t)]}~.
\end{equation}
Here, $n$ is a dimensionless azimuthal wavenumber;  $n = 0, 1, 2, 3, 4 ...$ describes  pinching, helical,
elliptical, triangular, rectangular, etc.\  distortions to the jet cross section. Formally there are an infinite number of solutions for each ``normal mode" $n$ with different radial structure \citep{H00}. In this paper we
only consider the low-$n$ modes, and argue that the elliptical mode dominates the structures seen in the M\,87 jet.

To derive the dispersion relation, we  assume  the jet has a ``top hat'' profile. We take density, pressure and velocity as  uniform across the jet, and assume that the external cocoon is in pressure balance with the jet, but
can otherwise have density and velocity different from those in the jet. In this situation, the normal modes are  governed by the wave dispersion relation, Equation (2) in \citet{H07} :
\begin{equation}
{ \zeta_j \over \xi_j} 
{ J_n' \left( \zeta_j r_j \right) \over J_n \left( \zeta_j r_j \right)}
=
{\zeta_c \over \xi_c} { H_n^{(1)'} \left( \zeta_c r_j \right) 
\over H_n^{(1)} \left( \zeta_c r_j \right)}
\label{DispRel}
\end{equation}
Here, $J_n$ and $H_n^{(1)}$ are Bessel and Hankel functions; the primes denote derivatives with respect to their arguments. The subscripts $j$ and $c$ refer to the jet and cocoon, and $r_j$ is the jet radius.  In the unmagnetized limit, the terms in Equation (\ref{DispRel}) are as follows.  The term $\xi$ is defined as $\xi \equiv \gamma^2 W \varpi^2$, with the plasma energy content $W = \rho ( 1 + \chi)$ defined in Equation (\ref{W_def}) and the specific enthalpy $\chi$ defined in Equation (\ref{chi_def}).  The factor $\varpi$ is defined by $\varpi^2 \equiv
\left( \omega - k u \right)^2$, for a flow speed $u = \beta c $ with a Lorentz factor $\gamma$. The term $\zeta$ is defined by $ \zeta^2 = \gamma^2 \ a^2 \left( \varpi^2 - \kappa^2 a^2 \right) $ where $a$ is the sound speed, given by Equation (\ref{sound_speed}), and $\kappa^2 = \left( k - \omega u/c^2 \right)^2$. Formally, this dispersion relation is for a cylinder.  \citet{H82} showed that a similar dispersion relation could be derived for conical expansion at much less than the Mach angle.  The local dispersion relation above delivers the same result if incremented along a conical jet using the appropriate instantaneous local conditions and spatial growth becomes power-law instead of exponential \citep{H87}.

For a given jet and cocoon flow speed, solutions of Equation (\ref{DispRel}) depend {\it only} on the thermal state of the jet and cocoon, which are described by $\chi_j$ and $\chi_c$ respectively.  At first glance, Equation (\ref{DispRel}) appears also to depend on the jet and cocoon densities, $\rho_j$ and $\rho_c$, through the $W_j$ and $W_c$ terms in the denominators of each side. However, because we assume pressure balance between the jet and the cocoon, the ratio $W_j / W_c$ in Equation (\ref{DispRel}) depends only on the ratio of the jet and cocoon enthalpies, $\chi_j/ \chi_c$.

The solution of  Equation (\ref{DispRel})  relates the wave frequency,
$\omega$, to a complex wavenumber, $k(\omega) = k_r(\omega) + i
k_i(\omega)$, for a given set of jet and cocoon parameters.  Formally
the solution to the dispersion relation provides an inverse complex
phase velocity as a function of $\omega$.  Thus, the real part of
$\omega / k$ determines both the wave speed, $v_w(\omega) = [\omega /
k(\omega)]_r$, and the wavelength, $\lambda_w(\omega) = 2 \pi
v_w(\omega)/\omega$.  Spatial growth or damping of the wave is
determined by the imaginary part of the inverse, i.e. by $ k_i / \omega$.  A negative value
for $k_i(\omega)$ indicates growth with $e$-folding length, $\ell
(\omega) \equiv |k_i(\omega)|^{-1}$.

In general, Equation (\ref{DispRel}) must be solved numerically, as we do in this paper.  However, the important propagation properties of this mode can be revealed by analytical approximations. The maximum growth rate for the elliptical surface mode occurs at a resonant frequency  (from Equation (16) in \citet{H07})
\begin{equation}
{\omega^{\ast} r_j \over a_{c}} \approx 
\frac{5\pi /4}{ \left[\left(1 - u_{c}/v^{\ast}_w\right)^2 -
\left(a_{c}/v^{\ast}_w - u_c a_{c}/c^2\right)^2\right]^{1/2}}~.
\label{resonant_freq}
\end{equation}
where the angular frequency $\omega$ refers to a perturbation propagating along the jet's surface at fixed cylindrical coordinate $\phi$.  For the elliptical mode this frequency and the accompanying wavelength correspond to a $180\arcdeg$ rotation of the elliptical cross section and high-pressure filaments and not a $360\arcdeg$ rotation of an individual high-pressure filament. The wave speed at resonance is  (from Equation (15) in \citet{H07})
\begin{equation}
v^{\ast}_{w}\approx \frac{\gamma _{j
}(\gamma _{ac}a_{c})u_{j}+\gamma _{c}(\gamma
_{aj}a_{j})u_{c}}{\gamma _{j}(\gamma _{ac}a_{c})+\gamma _{c}(\gamma
_{aj}a_{j})}~,
\label{resonant_wavespeed}
\end{equation}
where $\gamma _{a}\equiv (1-a^{2}/c^{2})^{-1/2}$ is the sonic Lorentz factor. The denominator on the right-hand side of Equation (\ref{resonant_freq}) shows a rapid increase in the resonant frequency as the jet becomes transonic. This means that the jet speed becomes comparable to the sound speed and $a_c/v^*_w$ increases with $v^*_w$ given by Equation (\ref{resonant_wavespeed}).  Note that the resonant frequency is inversely proportional to the jet radius and directly proportional to the cocoon sound speed provided the cocoon sound and flow speeds are small.

In the low-frequency limit ($\omega \ll  \omega^*$) the elliptical surface mode propagates and grows according to (from Equation (8) in \citet{H07})
\begin{equation}
\frac{\omega }{k}=\frac{\eta u_{j}+u_{c}}{1+\eta }\pm i\frac{\eta ^{1/2}}{
1+\eta }\left( u_{j}-u_{c}\right) ~ ,
\label{low_freq_DR}
\end{equation}
where $\eta \equiv \gamma_{j}^{2}W_{j}/ \gamma_{c}^{2}W_{c}$.  We see that $v_w = [ \omega / k(\omega)]_r$ can be large or small, according to whether  $\eta$ is large or small. In general, the resonant wave speed is faster than the low-frequency wave speed when $\eta$ is small and slower than the low-frequency wave speed when $\eta$ is large. Cocoon flow, $u_{c}$, can significantly affect the wave propagation speed and wavelength if $\eta < 1$.  From the solution to Equation (\ref{low_freq_DR}) we also estimate the growth length, $\ell(\omega) = | k_i(\omega)|^{-1}$.  When $u_j \gg u_c$,  which is the case in this paper, $\ell(\omega) / r_j  \simeq\eta^{1/2} u_j / \omega r_j$ for this low-frequency limit.

In the high-frequency limit ($\omega \gg \omega^*$)
\begin{equation}
\frac{\omega }{k}\approx \frac{u_{j}-a_{j}}{1-a_{j}u_{j}/c^{2}}~,
\end{equation}
for the unstable branch of the elliptical mode.  Here, a minimum possible wave speed occurs for $a_{j} = c/\sqrt 3$; for a relativistic jet with $u_{j} \sim c$ the intrinsic high frequency wave speed is $v_w > 0.63 c$. Since the condition $\eta \gg 1$, necessary for high wave speed in the low-frequency limit, typically requires $a_{j} \ll
c/\sqrt 3$, we see that the high-frequency wave speed exceeds the low-frequency wave speed even in this situation. In general, the wave speed at high frequencies exceeds the wave speed at frequencies below
resonance. In the high-frequency limit cocoon flow plays no role in the wave propagation speed.

\end{appendix}

\end{document}